\documentclass[pre,superscriptaddress,twocolumn,floatfix]{revtex4-2}

\usepackage{microtype}
\usepackage[utf8]{inputenx}
\usepackage{physics}                            
\usepackage{bm}									
\usepackage{enumerate}							
\usepackage{amsthm}								
\usepackage{comment}							
\usepackage[caption=false]{subfig}
\usepackage{amssymb}							
\usepackage{t1enc}								
\usepackage{mathtools}							
\usepackage{tensor}								
\usepackage{wrapfig}							
\usepackage{listings}							
\usepackage{xcolor}								
\usepackage{graphicx,color}
\usepackage{amsfonts,amsmath}
\usepackage{hyperref}
\hypersetup{colorlinks=true}

\usepackage[percent]{overpic}                   
\usepackage{pgfplots}                           
\usepackage{pgfplotstable}                      
\usepackage{scalefnt}                           
\usepgfplotslibrary{groupplots}
\usetikzlibrary{plotmarks}
\usetikzlibrary{external}   
\newcommand{\cor}[1]{\mathcal{#1}}									
\newcommand{\T}[1]{\text{#1}}										
\newcommand{\lgraf}{\left\lbrace}									
\newcommand{\rgraf}{\right\rbrace}									
\newcommand{\dslash}[1]{\frac{\dd[d]{#1}}{(2\pi)^d}}                
\newcommand{\n}{\nonumber}
\newcommand{\opunit}{\text{1}\kern-0.22em\text{l}}
\newcommand{\eg}{e.g.}
\newcommand{\ie}{i.e.}
\def \z {^{(0)}}
\def \o {^{(1)}}
\def \t {^{(2)}}

\newcommand{\pd}{\partial}

\newcommand{\0}{^{(0)}}
\newcommand{\1}{^{(1)}}
\newcommand{\2}{^{(2)}}

\newcommand{\Eq}[1]{Eq.\@~\eqref{#1}}
\newcommand{\ho}{h.o.}
\newcommand{\for}{\text{for}\,\,\,}
\definecolor{color1}{HTML}{D0B22B}

\begin{document}

\title{Non-equilibrium relaxation of a trapped particle in a near-critical Gaussian field}

\author{Davide Venturelli}
\affiliation{SISSA -- International School for Advanced Studies and INFN, via Bonomea 265, 34136 Trieste, Italy}
\author{Francesco Ferraro}
\affiliation{Alumnus, Physics Department, University of Trento, via Sommarive, 14 I-38123 Trento, Italy}
\author{Andrea Gambassi}
\affiliation{SISSA -- International School for Advanced Studies and INFN, via Bonomea 265, 34136 Trieste, Italy}

\begin{abstract}
We study the non-equilibrium relaxational dynamics of a probe particle linearly coupled to a thermally fluctuating scalar field and subject to a harmonic potential, which provides a cartoon for an optically trapped colloid immersed in a fluid close to its bulk critical point. The average position of the particle initially displaced from the position of mechanical equilibrium is shown to feature long-time algebraic tails as the critical point of the field is approached, the universal exponents of which are determined in arbitrary spatial dimensions. As expected, this behavior cannot be captured by adiabatic approaches which assume fast field relaxation. The predictions of the analytic, perturbative approach are qualitatively confirmed by numerical simulations.
\end{abstract}

\maketitle
\section{Introduction}
Studying the motion of colloidal particles in contact with thermally fluctuating environments provides a tool to probe the rheological properties of soft-matter systems \cite{squires2005simple, zia2013stress}. While past studies have mostly focused on the behavior of tracer particles passively carried by a fluctuating solvent, in recent years increasing attention has been paid to cases in which the particle and the solvent affect each other dynamically \cite{demery2010, demery2010-2, demery2011, demery2013, demerypath, Dean_2011, fuji1, fuji2}. 
Particularly interesting is the case in which the medium is a fluid near a critical point, thus displaying long-range spatial correlations and long relaxation times. Objects immersed in near-critical fluids are known to experience fluctuation-induced forces \cite{kardar99,GambassiCCF, Maciolek} such as the critical Casimir force, \ie, the thermal analog of the celebrated effect in quantum electrodynamics \cite{Casimir}. While equilibrium field-mediated effects have long since been explored, the dynamical behavior of these systems has rarely been addressed in the literature. Here we wish to start filling this gap by analyzing a simple setup and by predicting the dynamics of quantities which are easily accessible in experiments. The paradigm we have in mind is that of a near-critical fluid such as a binary liquid mixture \cite{casimirColloids,energytransfer}, in which a colloidal particle is trapped by optical tweezers, and we measure the average and the correlation functions of its position.

In this work we study the non-equilibrium dynamics of a probe particle in contact with a fluctuating medium close to the bulk critical point of a continuous phase transition, and trapped in a harmonic potential. The medium is modeled as a scalar order parameter $\phi(\vb{x})$ subject to a dissipative or conserved relaxational dynamics within the Gaussian approximation (model A and B, respectively, in the classification of Ref.~\cite{halperin}), while the probe represents an overdamped colloidal particle interacting with the scalar field via a translationally invariant linear coupling. Because of this coupling, the particle and the field affect each other dynamically along their stochastic evolution, in such a way that detailed balance is fulfilled at all times. Despite its simplicity, this minimal model already displays nonlinear and non-Markovian effects in the resulting dynamics of the colloid, which make analytical predictions difficult beyond perturbation theory.

Here we focus our attention on the dynamics of the probe particle and we study how it is affected by the presence of the field. A recent work \cite{wellGauss} investigated the auto-correlation function of the particle fluctuating in the harmonic trap in contact with a Gaussian field with conserved dynamics (model B), a problem which was tackled within the weak-coupling approximation. In particular, this proved the emergence of algebraic tails at long times superimposed to the usual exponential decay of the auto-correlation, the exponent of which depends only on the spatial dimensionality of the system. These results do not depend on the details of the chosen interaction potential between the colloid and the medium, provided that it is linear and translationally invariant.

A similar setup was analyzed in Ref.~\cite{gross}, where the steady-state and effective dynamics of a colloid in contact with a critical Gaussian field were investigated in the presence of spatial confinement for the field. In the case of a linear coupling between the fluctuating field and the colloid, an effective Fokker-Planck equation was obtained under the assumption of rapid relaxation of the field for each position of the particle. This allowed the adiabatic elimination of the field degrees of freedom, given by its eigenmodes in a finite box subject to certain boundary conditions, from the coupled equations of motion of the system.

Our aim here is to analyze the relaxation of the particle after it is released far from its position of mechanical equilibrium in the harmonic trap. Within a weak-coupling expansion, we first show that the average position of the colloid itself displays an algebraic behavior at long times, and we relate its decay exponents to those of the auto-correlation function of the position of the colloid in view of the fluctuation-dissipation theorem. We then interpret these dynamical exponents only in terms of the spatial dimensionality of the system and the dynamical critical exponent $z$ of the field. Our analysis additionally reveals a transient algebraic behavior which is entirely due to the nonlinearity of the effective particle dynamics, and which is therefore out of the reach of linear response theory. We test our perturbative, analytical predictions against numerical simulations of the complete system in order to exclude the possibility that higher-order corrections in the coupling constant $\lambda$ become increasingly relevant at long times; this way we prove that the qualitative features of our analytical predictions, based on a perturbative expansion in $\lambda$, remain valid beyond perturbation theory.

In the same spirit as Ref.~\cite{gross}, we then derive an effective Fokker-Planck equation for the motion of the colloid in the adiabatic limit by integrating out the field degrees of freedom, which are a continuum of variables in the bulk. We use this effective equation to study again the problem of relaxation towards equilibrium and we investigate on the possible matching between the perturbative and the adiabatic predictions; this allows us to locate precisely the point at which the adiabatic approximation breaks down. In particular we find, as expected, that the latter fails close to criticality and even far from criticality when the field dynamics is conserved.

The rest of the presentation is organized as follows. In Section \ref{par:Model} we introduce the model and the notation. In Section \ref{par:Relaxation} we study the problem of relaxation towards equilibrium using a weak-coupling expansion, while in Section \ref{par:Adiabatic} we consider the same problem but within the adiabatic approximation (the details of which are presented in Appendix \ref{par:adiabatic}); we then compare the two approaches and thus determine the limits of validity of the adiabatic approximation. In Section \ref{par:numerical} we present numerical simulations supporting our analytical predictions and we use them to provide a qualitative description of the relaxation beyond the linear regime. We finally summarize our results in Section \ref{par:conclusion}.

\section{The model} 
\label{par:Model}
\begin{figure}
\centering
\begin{overpic}
    [width=\columnwidth]{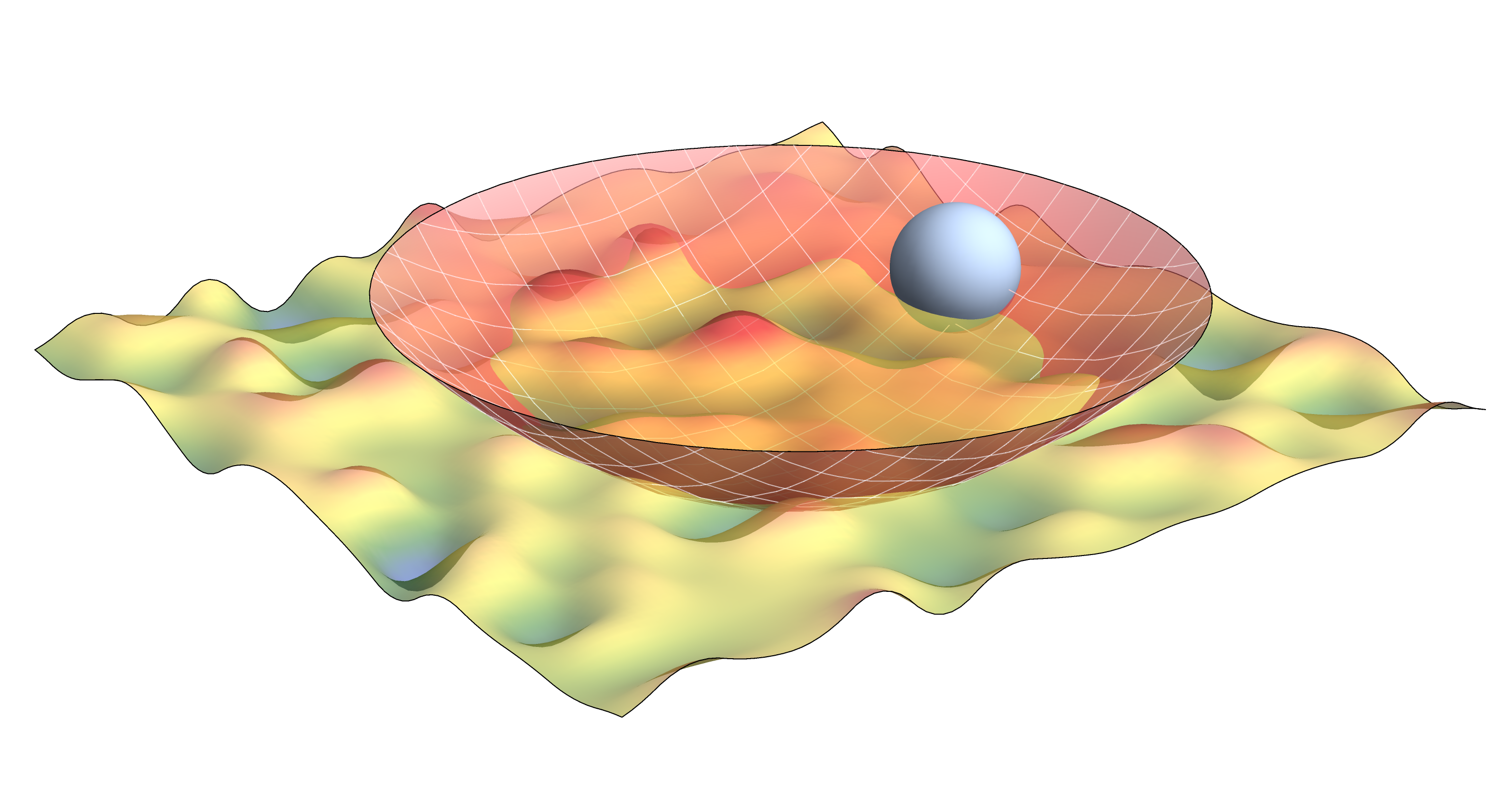}
    \put(36,13){\large $\displaystyle\phi(\vb{x})$}
    \put(69,34){\large \color{white} $\displaystyle \vb{X}$}
\end{overpic}
\caption{Pictorial representation of the model: a colloidal particle is in contact with a fluctuating scalar field $\phi(\vb{x})$ and trapped by a harmonic potential.}
\label{fig:setup}
\end{figure}

The system composed by the particle and the field is described by the Hamiltonian \cite{wellGauss}
\begin{align}
    \cor{H}=&    \int \dd[d]{\vb{x}}
    \left[\frac{1}{2}(\grad\phi)^2+\frac{r}{2}\phi^2 \right]
    + \frac{k}{2}\vb{X}^2 \n \\
    & - \lambda
    \int \dd[d]{\vb{x}}
    \phi(\vb{x})V(\vb{x}-\vb{X}) \; ,
\label{eq:hamiltonian}
\end{align}
where $\phi$ is a scalar Gaussian field in $d$ spatial dimensions, and the $d$-dimensional vector $\vb{X}$ denotes the position of a reference point on the probe particle, \eg, its center. The constant $k$ sets the strength of the harmonic potential in which the particle is trapped, while $r\geq 0$ is a measure of the deviation from criticality and controls the correlation length $\xi = r^{-1/2}$ of the fluctuations of the field at equilibrium. The system is schematically represented in Fig.~\ref{fig:setup}.

The coupling between the particle and the field is linear and translational invariant: this may physically model, for example, a colloid displaying a preferential adsorption towards one of the two components of a binary mixture. 
The interaction potential $V(\vb{x})$ is a function which models the shape of the colloidal particle, in the sense that the field interacts with the colloid within its spatial extent, determined by the support of $V(\vb{x})$. For spherically symmetric tracers, it can be chosen as $V(\vb{x})=\delta(\vb{x})$ in the case of a point-like colloid, while we will consider a Gaussian form of $V(\vb{x})$ with variance $R$ whenever we need to keep track of the particle size. We choose $V(\vb{x})$ to be normalized so that its integral over all space is equal to unity; this way the strength of the interaction is set only by the coupling constant $\lambda$. If $\lambda$ and $V(\vb{x})$ in Eq.~\eqref{eq:hamiltonian} are chosen to be positive, then configurations are favored in which the field $\phi$ is enhanced and assumes preferentially positive values in the vicinity of the colloidal particle.

Adopting the minimal model in Eq.~\eqref{eq:hamiltonian} is physically motivated as follows. Upon approaching the critical point, spatial correlations in the medium and consequently the characteristic timescale of its dynamics grow arbitrarily large; the system thus displays universal features, increasingly independent of its microscopic details, and a minimal description of the medium in terms of a suitably-chosen coarse-grained order parameter is sufficient as long as one is interested in its long-range and long-time behaviour. Moreover, a mesoscopic colloidal particle evolves on much longer timescales than the microscopic degrees of freedom of the medium, due to their difference in size. It is then expected that the particle coordinate and the order parameter are the slow degrees of freedom, while all other degrees of freedom effectively generate a random forcing. In this model, we consider a simple scalar order parameter $\phi(\vb{x},t)$, while we neglect hydrodynamics effects and other slow variables which should be taken into account when describing real fluids or binary liquid mixtures \cite{halperin}.

We assume a purely relaxational dynamics for the field,
\begin{align}
        &\partial_t\phi(\vb{x},t)= -D (i \grad)^\alpha \fdv{\cor{H}}{\phi(\vb{x},t)} + \zeta(\vb{x},t)     \label{eq:field}\\
        &= -D(i \grad)^\alpha \left[(r-\nabla^2)\phi(\vb{x},t)-\lambda V(\vb{x}-\vb{X}) \right] + \zeta(\vb{x},t) \; .\n
\end{align}
Here $\alpha=0$ for a non-conserved dynamics of the order parameter $\phi$, while $\alpha=2$ if $\phi$ is subject to local conservation during the evolution, in the sense that Eq.~\eqref{eq:field} can then be cast in the form $\partial_t\phi(\vb{x},t)=-\div \vb{J}(\vb{x},t)$ with a suitably chosen current $\vb{J}(\vb{x},t)$. These two choices of $\alpha$ correspond to model A and model B in the classification of Ref.~\cite{halperin}, in which we neglect the self-interaction term $\sim \phi^4$, \ie, within the Gaussian approximation. Finally, $\zeta(\vb{x},t)$ is a white Gaussian noise with zero mean and variance
\begin{equation}
    \expval*{\zeta(\vb{x},t)\zeta(\vb{x}',t')}= 2DT (i \grad)^\alpha \delta^d(\vb{x}-\vb{x}')\delta(t-t')\; ,
\end{equation}
with $D$ and $T$ denoting, respectively, the mobility (or diffusivity) of the field and the temperature of the environment.

The dynamics of the probe particle is described by the overdamped Langevin equation
\begin{align}
        \dot{\vb{X}}(t)&= -\nu \grad_X \cor{H}   + \bm{\xi}(t) \n \\
        &=-\nu k\vb{X} + \nu\lambda \vb{f} + \bm{\xi}(t) \; ,
    \label{eq:particle}
\end{align}
where the force $\vb{f}$ acting on the particle is given by the gradient of the interaction energy \footnote{We adopt here and in the following the Fourier convention $ f(\vb{x}) = \int \frac{\dd[d]{q}}{(2\pi)^d} e^{i\vb{q}\cdot \vb{x}} f_{\vb{q}}$.}
\begin{align}
        \vb{f}(\vb{X},\phi;t) &\equiv \grad_{\vb{X}} \int \dd[d]{x} \phi(\vb{x},t)V(\vb{x}-\vb{X}(t)) \n \\
        &= \int \dslash{q} i \vb{q} \phi_q(t) V_{-q} e^{i\vb{q}\cdot \vb{X}(t)} \; .
    \label{eq:f}
\end{align}
The particle and the field are assumed to be in contact with the same thermal bath at temperature $T$, so that $\bm{\xi}(t)$ is also a white Gaussian noise with zero mean and variance
\begin{equation}
    \expval*{\xi_{i}(t) \xi_{j}(t') } = 2\nu T \delta_{ij}\delta(t-t') \; ,
    \label{eq:part_noise}
\end{equation}
where $\nu$ is the mobility of the probe.

The Langevin equation for the field in Fourier space reads
\begin{equation}
\dot{\phi}_q = -\alpha_q \phi_q +D\lambda q^\alpha V_q e^{-i\vb{q}\cdot \vb{X}}+ \zeta_q \; ,
\label{eq:fieldFourier}
\end{equation}
where we introduced $\alpha_q\equiv Dq^\alpha(q^2+r)$, while \footnote{The common practice of bringing around $(2\pi)^d$ factors in these formulas can be simply avoided by defining the delta distribution in Fourier space as $\int \dslash{q} \delta^d(q)=1$. We adopt here this definition.}
\begin{equation}
    \expval*{\zeta_q(t)\zeta_{q'}(t')}= 2DTq^\alpha \delta^d(q+q')\delta(t-t') \; .
    \label{eq:field_noise}
\end{equation}
It must be noted that an unbounded growth of the zero mode $\phi_{q=0}$ is implied by Eq.~\eqref{eq:fieldFourier} for model A dynamics when $r=0$. While this has no consequence on the particle dynamics (see Eq.~\eqref{eq:f}), in a more realistic system one would need to counteract this growth by adding a suitable chemical potential.

Upon switching off the coupling between the particle and the field, \ie, setting $\lambda = 0$, the two stochastic processes are non-interacting and their solution is summarized in Appendix \ref{par:correlators}. They are characterized by the (inverse) relaxation timescales
\begin{align}
    \tau_X^{-1}&=\nu k \equiv \gamma \; , \label{eq:tau_x}\\
    \tau_\phi^{-1}(q) &= \alpha_q = Dq^\alpha(q^2+r) \label{eq:tau_phi} \; .
\end{align}
In particular, the relaxation time $\tau_\phi(q\sim 0)$ for the long-wavelength modes of the field may become arbitrarily large for model A dynamics at $r=0$. The same happens for model B dynamics for generic values of $r$, \ie, also off-criticality, due to the presence of the conservation law for which $\tau_\phi^{-1}(q\rightarrow 0)=0$. These long-wavelength modes are always present in the bulk, while they are cut-off in a confined geometry such as that considered in Ref.~\cite{gross}.

Since the dynamics in Eqs.~\eqref{eq:field} and \eqref{eq:particle} satisfies detailed balance, the joint equilibrium distribution of the field and the particle is the canonical one,
\begin{align}
    P_\T{eq}[\phi,\vb{X}] \propto \exp(-\beta \cor{H}[\phi,\vb{X}]) \; ,
\end{align}
where $\beta = 1/T$. Accordingly, the equilibrium distribution $P_\T{eq}(\vb{X})$ of the colloid is found by marginalizing $P_\T{eq}[\phi,\vb{X}]$ as
\begin{align}
    P_\T{eq}(\vb{X}) \propto \int\cor{D}\phi \, e^{-\beta \cor{H}[\phi,\vb{X}]} \; .
\end{align}
We show in Appendix \ref{par:eqcolloid} that $P_\T{eq}(X)$ is actually not affected by the presence of the field, and one still finds $P_\T{eq}(X) \propto \exp(-\beta k X^2/2)$. The argument we invoke is completely general: it relies neither on the linearity of the coupling nor on the choice of a free field theory, and not even on the use of a quadratic particle potential. The only requirement is that the dynamics occurs in the bulk (\ie, there must be no boundaries) and that the coupling between the field and the particle is translationally invariant. We emphasize that the equilibrium distribution of the colloid would indeed depend on the kind of coupling and boundary conditions if we had considered a system in a confined geometry, thus breaking translational invariance \cite{gross}. Moreover, even in the bulk considered here, the marginal equilibrium distribution $P_\T{eq}[\phi]$ of the field alone \textit{does} indeed get modified by the presence of $\vb{X}$.

Non-trivial aspects of the  field-particle interaction can nonetheless be deduced by looking at the dynamical properties of the probe. In this work we therefore set out to predict the dynamics of the average value of the position of the colloidal particle as it relaxes towards the center of the harmonic trap, being initially displaced from the position of mechanical equilibrium corresponding to $\vb{X}=0$.

\section{Weak-coupling approximation}
\label{par:Relaxation}

\begin{figure*}
\centering
\subfloat{
  \centering
  \includegraphics[height = 0.29\textwidth]{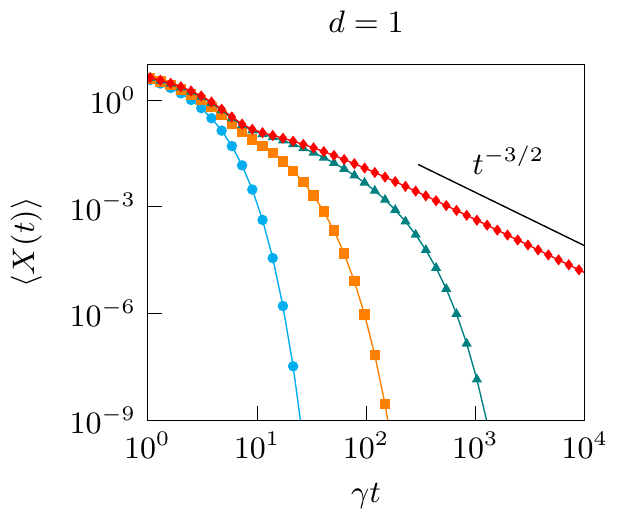}
}
\subfloat{
  \centering
  \includegraphics[height = 0.29\textwidth]{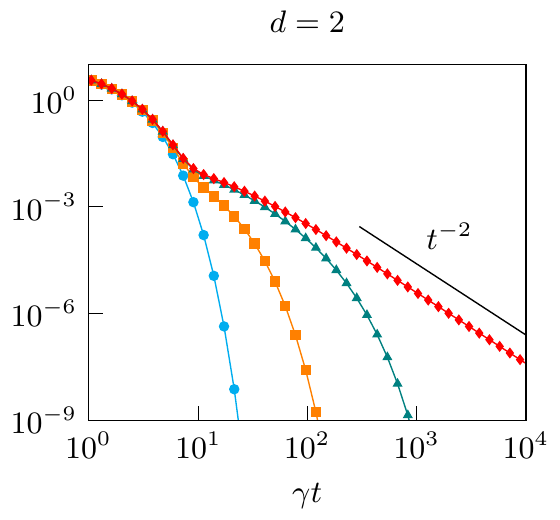}
}
\subfloat{
  \centering
  \includegraphics[height = 0.29\textwidth]{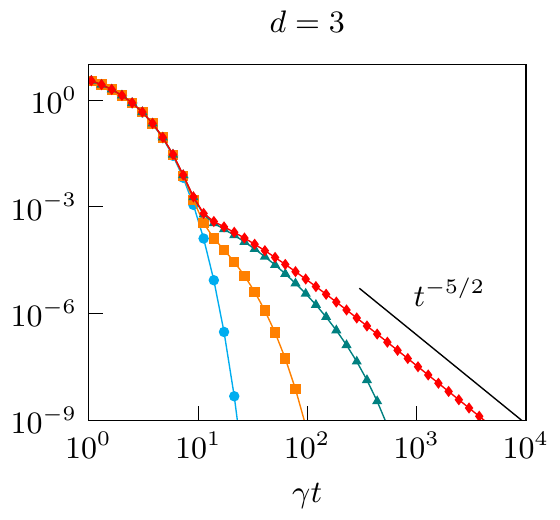}
}\\
\includegraphics{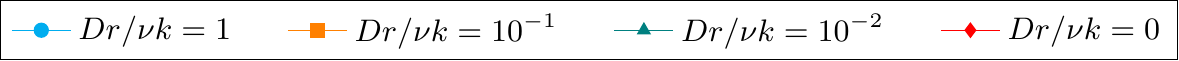}
\caption{Evolution of the average position $\expval{X(t)}$ of a harmonically trapped particle initially released out of equilibrium and coupled to a field evolving with model A dynamics, for various decreasing distances $r$ from the critical point. The plots show the analytical prediction up to $\order{\lambda^2}$. The points are obtained from the numerical integration of Eq.~\eqref{eq:X2_first} and they are joined by a linear interpolation to guide the eye. The exponents observed at criticality, $r=0$, agree with those predicted in  Eq.~\eqref{eq:relax_d_modelA}. In these plots $\lambda=1$, $\nu=10$, $k=0.1$, $T=1$, $D=1$, $X_0=10$, and the interaction potential is Gaussian with $R=1$.}
\label{fig:relax_A}
\end{figure*}
\begin{figure*}
\centering
\subfloat{
  \centering
  \includegraphics[height = 0.29\textwidth]{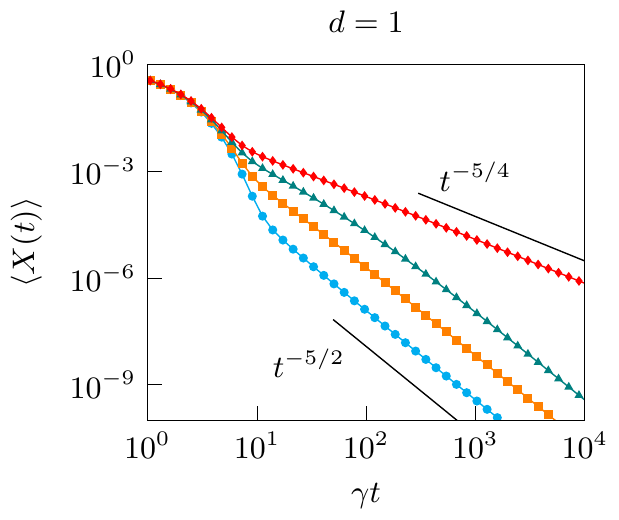}
}
\subfloat{
  \centering
  \includegraphics[height = 0.29\textwidth]{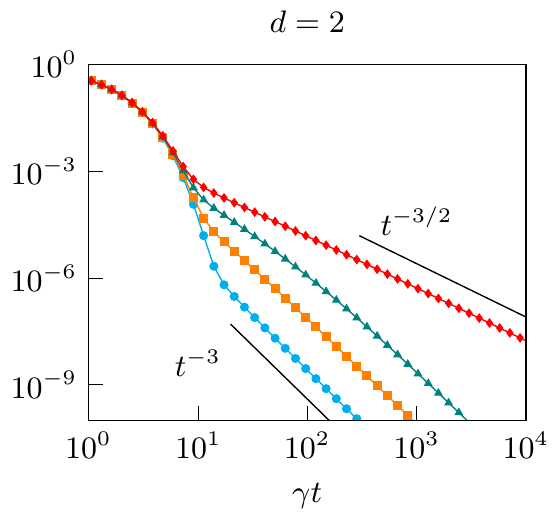}
}
\subfloat{
  \centering
  \includegraphics[height = 0.29\textwidth]{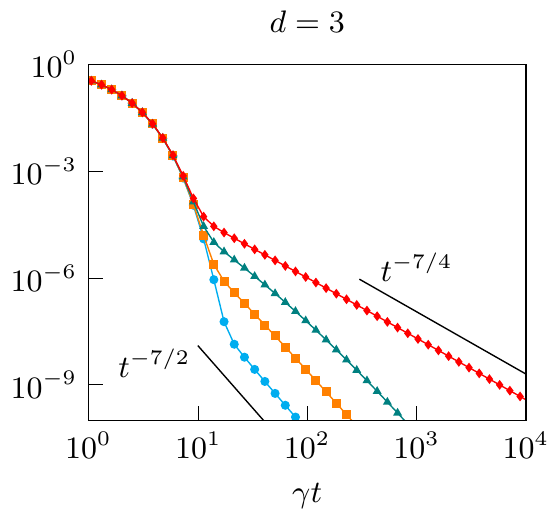}
}\\
\includegraphics{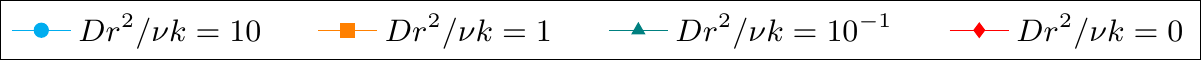}
\caption{Evolution of the average position $\expval{X(t)}$ of a harmonically trapped particle initially released out of equilibrium and coupled to a field evolving with model B dynamics, for various decreasing distances $r$ from the critical point. The plots show the analytical prediction up to $\order{\lambda^2}$. The points are obtained from the numerical integration of Eq.~\eqref{eq:X2_first} and they are joined by a linear interpolation to guide the eye. The decay exponents agree with those predicted in \Eq{eq:relax_d}. In these plots $\lambda$, $\nu$, $k$, $T$, $D$ and $X_0$ are set to unity and the interaction potential is Gaussian with $R=1$.}
\label{fig:relax_B}
\end{figure*}

The coupled nonlinear equations \eqref{eq:field} and \eqref{eq:particle} for the dynamics of the particle and the field are not exactly solvable and we therefore resort to a perturbative expansion in the coupling strength $\lambda$, computing the relevant observables at the lowest nontrivial order in this parameter \cite{wellGauss}. It must be noted that $\lambda$ is not dimensionless: dimensional analysis of the Hamiltonian in Eq.~\eqref{eq:hamiltonian} gives $[\phi]= d/2-1$ and accordingly $[\lambda]= 1-d/2$ for the dimensions $[\phi]$ and $[\lambda]$ of the field and the coupling respectively, in units of inverse length. 

We consider the following formal expansions of the field and of the coordinates of the particle:
\begin{align}
    \phi(\vb{x},t) &= \sum_{n=0}^\infty \lambda^n\phi^{(n)}(\vb{x},t) \; , \n \\
    \vb{X}(t) &= \sum_{n=0}^\infty \lambda^n \vb{X}^{(n)}(t) \; . 
    \label{eq:Xseries}
\end{align}
These can be inserted into Eq.~\eqref{eq:particle} for the particle to get, order by order in the coupling $\lambda$,
\begin{align}
    \dot{\vb{X}}\z(t) &= -\nu k \vb{X}\z(t) + \bm{\xi}(t) \; , \label{eq:perturb}\\
    \dot{\vb{X}}^{(n)}(t) &= -\nu k \vb{X}^{(n)}(t) + \nu \vb{f}^{(n-1)}(t) \; ,
\end{align}
where we introduced
\begin{equation}
    \vb{f}^{(n)}(t) \equiv \frac{1}{n!}\eval{\dv[n]{\lambda}}_{\lambda=0} \vb{f}(t) \; .
\end{equation}

At $\order{\lambda^0}$, Eq.~\eqref{eq:perturb} is solved by the Ornstein-Uhlenbeck process, recalled in Appendix \ref{par:correlators}. The higher-order corrections $\vb{X}^{(n)}$ can be formally expressed as
\begin{equation}
    \vb{X}^{(n)}(t) = \nu \int_{t_0}^t \dd{s} e^{-\gamma (t-s)} \vb{f}^{(n-1)}(s) \; ,
\end{equation}
where $t_0$ is the time at which the initial condition $\vb{X}\z(t=t_0)=\vb{X}_0$ is imposed.
Similarly, the Langevin equation~\eqref{eq:fieldFourier} for the field in Fourier space renders
\begin{align}
    \pd_t\phi^{(0)}_q(t) 
        &= - \alpha_q \phi^{(0)}_q(t) + \zeta_q(t), 
    \label{eq:phi0} \\
    \pd_t\phi^{(n)}_q(t) 
        &= - \alpha_q \phi^{(n)}_q(t) +\frac{D q^\alpha V_q}{(n-1)!} \eval{\dv[n-1]{\lambda}}_{\lambda=0} e^{-i\vb{q}\cdot \vb{X}} \; ,
    \label{eq:phin}
\end{align}
where $\vb{X}$ on the r.h.s. of Eq.~\eqref{eq:phin} is written in powers of $\lambda$ as in Eq.~\eqref{eq:Xseries}. The function $V_q$ is the Fourier transform of the interaction potential $V(\vb{x})$, and it only depends on $\abs{\vb{q}}$ if we take $V(\vb{x})$ to be isotropic, \ie, a function of $|\vb{x}|$. The properties of the uncoupled field $\phi^{(0)}_q(t)$ are discussed in Appendix \ref{par:correlators}, while the equation of motion of the field at $\order{\lambda}$ can be formally solved as
\begin{equation}
    \phi_q \o (s) = Dq^\alpha V_q  \int_{t_0}^s \dd{\tau} e^{-\alpha_q (s-\tau)} e^{-i\vb{q}\cdot \vb{X}\z (\tau)} \; .
    \label{eq:phi1}
\end{equation}
If we assumed the field to be initially in thermal equilibrium \textit{in contact} with the colloid, then a second term accounting for the initial condition of the field would appear in Eq.~\eqref{eq:phi1} in the form $G_q(t-t_0) \phi_q \o (t_0)$, where the function $G_q(\tau)$ is the free-field propagator defined in Appendix \ref{par:freefield}. However, such a term turns out \textit{a posteriori} to be irrelevant for what concerns the long-time properties of the tracer particle. For the sake of simplicity, we will assume instead that the initial condition of the field $\phi_q(t_0)$ is extracted from its stationary distribution reached \textit{before} the colloid is put in contact with the field.

\subsection{Perturbative corrections to the position}
\label{par:pert1p}
The average particle position is given by
\begin{equation}
    \expval*{\vb{X}(t)} = \expval*{\vb{X}\z (t)} + \lambda^2 \expval*{\vb{X}\t (t)} + \order{\lambda^4} \; ,
    \label{eq:x_perturbative}
\end{equation}
because one can argue on the basis of the invariance of the equations of motion under $\lambda\leftrightarrow -\lambda$, $\phi\leftrightarrow -\phi$ that $\expval*{\vb{X}\o (t)} = 0$ \cite{wellGauss}. The first nontrivial term is thus of $\order{\lambda^2}$ and it can be computed starting from
\begin{align}
    \vb{f}\z(s_1) =&  \int \frac{\dd[d]{q}}{(2\pi)^d} i \vb{q} V_{-q}\phi_q \z (s_1) e^{i\vb{q}\cdot \vb{X}\z (s_1)}\; , \n\\ 
    \vb{f}\o(s_2) =&  \int \frac{\dd[d]{q}}{(2\pi)^d} i \vb{q} V_{-q} e^{i\vb{q}\cdot \vb{X}\z (s_2)} \Big[ \phi_q \o (s_2) \n \\
    & + i\vb{q}\cdot \vb{X}\o (s_2)\phi_q \z (s_2)  \Big] \; ,
\end{align}
while bearing in mind that, as we take the expectation values over the realizations of the noises $\zeta_q(t)$ and $\bm{\xi}(t)$,
\begin{align}
    &\expval*{\phi_q \z (s_2)\phi_q \z (s_1) e^{i \vb{q} \cdot [\vb{X}\z(s_2)- \vb{X}\z(s_1)]} } \n \\[2mm]
    &= \expval*{\phi_q \z (s_2)\phi_q \z (s_1)} \expval*{  e^{i \vb{q} \cdot [\vb{X}\z(s_2)- \vb{X}\z(s_1)]}} \; ,
\end{align}
because at $\order{\lambda^0}$ the two processes $\phi_q \z(t)$ and $\vb{X}\z(t)$ are independent. Using $V_{-q}=V^*_{q}$ because $V(\vb{x})$ is real, we find

\begin{widetext}
\begin{align}
    \expval*{X_j\t (t)}  = \nu \int \frac{\dd[d]{q}}{(2\pi)^d} i q_j  \abs{V_q}^2 \int_{t_0}^t \dd{s_2} e^{-\gamma (t-s_2)} \int_{t_0}^{s_2} \dd{s_1}\left[ \chi_q(s_1,s_2) +\nu q^2e^{-\gamma (s_2-s_1)} C_q(s_1,s_2) \right] Q_q(s_1,s_2) \; .
    \label{eq:X2_first}
\end{align}
\end{widetext}

We have introduced
\begin{align}
    &Q_q(s_1,s_2) \equiv \expval*{ e^{i \vb{q} \cdot [\vb{X}^{(0)}(s_2)- \vb{X}^{(0)}(s_1)]} } \; , \label{eq:Qq_def} \\[2mm]
    &\expval*{\phi_q\z(s_1)\phi_p\z(s_2)} = \delta^d(p+q)C_q(s_1,s_2) \; ,
\end{align}
where the averages are taken over the non-interacting processes with $\lambda=0$; they are computed by standard methods in Appendix \ref{par:correlators}. The functions $C_q(s_1,s_2)$ and $\chi_q(s_1,s_2)$ are, respectively, the stationary correlator of the Gaussian field in the absence of the particle and its dynamical susceptibility (see Ref.~\cite{Tauber} and Appendix \ref{par:freefield}), given by
\begin{align}
    &C_q(s_1,s_2) = C_q(s_2-s_1)= \frac{T}{q^2+r} e^{-\alpha_q\abs{s_2-s_1}} \; , \label{eq:Cqfield} \\
    &\chi_q(s_1,s_2) = \chi_q(s_2-s_1) = Dq^\alpha e^{-\alpha_q (s_2-s_1)} \theta(s_2-s_1)
    \label{eq:field-susc} \; ,
\end{align}
where $\theta(x)$ is the Heaviside distribution.

We first specialize Eq.~\eqref{eq:X2_first} to the case of a particle leaving at time $t=t_0$ the initial position $\vb{X}(t_0)=\vb{X}_0\neq 0$ (the above expression for $\expval*{X_j\t (t)}$ would remain valid if the initial condition $\vb{X}\z(t_0)$ were drawn instead from a random distribution). The asymptotic behavior of the resulting $\vb{X}(t)$ at long times is then examined in Appendix \ref{par:appaverageposition}, where we consider the general case in which $V_q \sim q^n$. Although we have assumed $V(\vb{x})$ to be normalized to unity in real space (hence $V_{q=0}=1$), this may model the case in which the colloid is linearly coupled to the $n$-th (even) derivative of the field via an interaction term of the form
\begin{equation}
    \cor{H}_\T{int}=  -\lambda\int \dd[d]{\vb{x}} V(\vb{x}-\vb{X}) \grad^n \phi(\vb{x})
    \label{eq:derivativeinteraction}
\end{equation}
in the Hamiltonian in Eq.~\eqref{eq:hamiltonian}. Below we summarize the main results of this analysis. In order to lighten the notation, we will often omit the suffix $j$ from $\expval*{X_j(t)}$, since its only non-zero component is the one along the initial displacement $\vb{X}_0$.

\begin{figure}[b]
    \centering
    \includegraphics[width=\columnwidth]{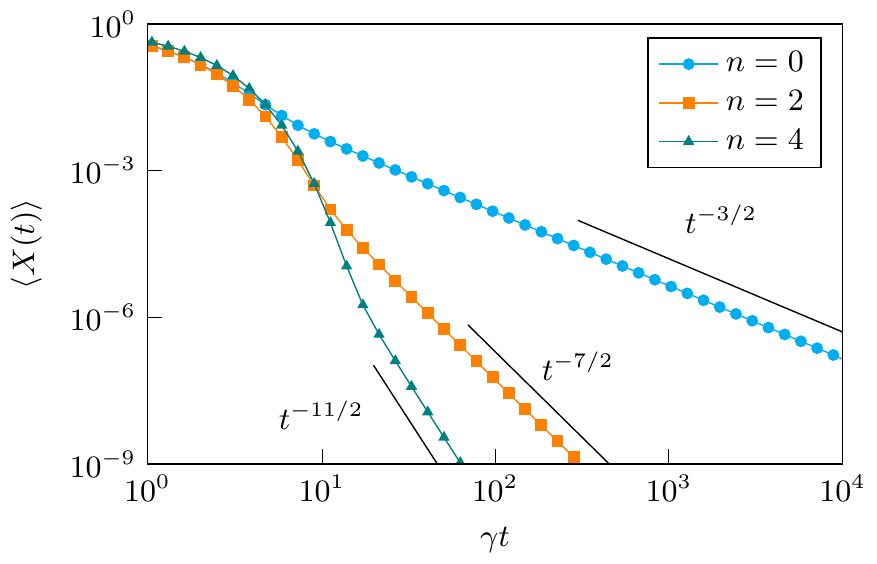}
    \caption{Average particle position $\expval*{X(t)}$ in the presence of a coupling to the $n$-th derivative of the field, as in Eq.~\eqref{eq:derivativeinteraction}. The plot shows the case of model A in $d=1$, and the observed decay exponents agree with those predicted in Eq.~\eqref{eq:general_critical} for $z=2$. In this plot $\lambda$, $\nu$, $k$, $T$, $D$ and $X_0$ are set to unity and the interaction potential is Gaussian with $R=1$.}
    \label{fig:couplingtype}
\end{figure}

\subsection{Long-time behavior of the position}
\label{par:longtime}
By direct inspection of Eq.~\eqref{eq:X2_first} in the case of model A and B dynamics (see Appendix \ref{par:appaverageposition}), we find the long-time asymptotics of the mean particle position $\expval*{X(t)}$ to be given in model A by
\begin{equation}
    \expval*{X(t)} \sim
        \begin{cases}
            e^{-\gamma t}  & \T{for}\;\; r>\gamma/D \; , \\
            e^{-Dr t}    & \T{for}\;\; r< \gamma/D \; , \\
            t^{-\left(1+d/2\right)} & \T{for}\;\;  r=0 \; ,
        \end{cases}
    \label{eq:relax_d_modelA}
\end{equation}
and in model B by
\begin{equation}
    \expval*{X (t)} \sim
    \begin{cases}
        t^{-\left(2+d/2 \right)} & \T{for}\;\;  r>0 \; , \\
        t^{-\left(1+d/4 \right)} & \T{for}\;\; r=0    \; .
    \end{cases}
    \label{eq:relax_d}
\end{equation}
These results have a clear physical interpretation: the long-time dynamics of the particle is practically determined by the slowest timescale characterizing the system. The two competing timescales are given by $\tau_X$ and $\tau_\phi$ in Eqs.~\eqref{eq:tau_x} and \eqref{eq:tau_phi}, where we set $q=0$ in the latter in order to account for the longest wavelength mode, which is infinite in the bulk. Consider first the case of model A dynamics, for which $\tau_\phi^{-1}(q=0)=Dr$. Sufficiently away from the critical point, \ie, for large $r$, where the field evolves more rapidly than the particle, the motion of the latter is essentially unaffected. Upon approaching criticality, \ie, by reducing the value of $r$ towards $0$, the dynamics of the field becomes instead increasingly slower and eventually it becomes the longest timescale: this determines the change in the rate of exponential decay observed in Eq.~\eqref{eq:relax_d_modelA}. Finally at criticality, $r=0$, the divergence of the timescale characterizing the field dynamics induces correspondingly a scale-free behavior of the tracer particle. In model B, on the other hand, $\tau_\phi(q=0)$ is formally infinite even away from criticality, due to the presence of a conservation law: as a result, the dynamics of the tracer particle is always controlled by the field for any value of the parameter $r$.

\begin{figure}
    \centering
    \includegraphics[width=\columnwidth]{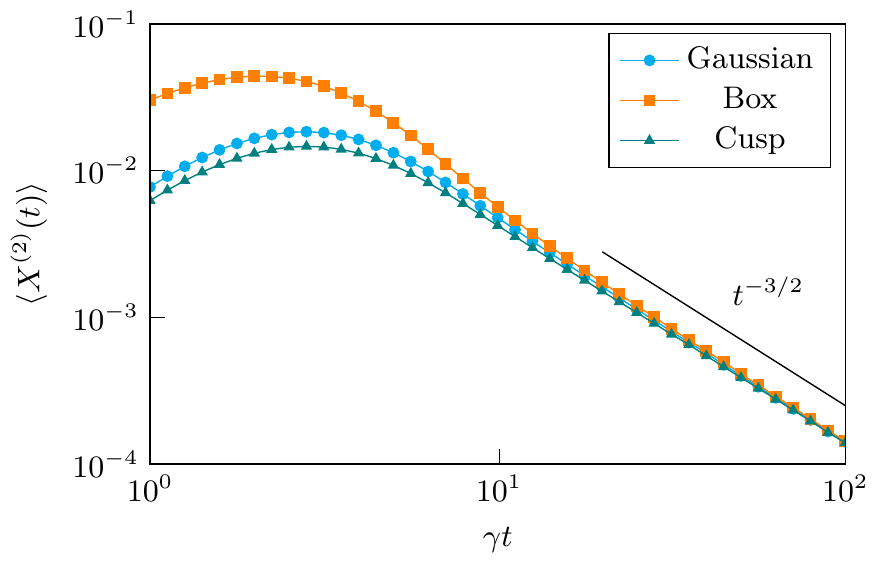}
    \caption{Independence of the long-time behaviour of the average position of the colloid from the particular choice of the interaction potential $V(\vb{x})$. The plot shows the correction $\expval*{X^{(2)}(t)}$ for model A in $d=1$ (the leading order exponential term is irrelevant at long times). The exponent of the algebraic decay is not affected by the specific form of $V(\vb{x})$, but only by the behavior of its Fourier transform $V_q$ for $q \rightarrow 0$, in agreement with the asymptotic expression in Eq.~\eqref{eq:mean_asymptotic}. On the contrary, the short-time behavior is sensitive to the particular choice of $V(\vb{x})$. The forms of the interaction potential reported here are Gaussian $V_q=\exp(-R^2 q^2/2)$, box $V_q=\textrm{sinc}(R q/2)$, cusp $V_q=1/(1+R^2q^2)$. Here $R$ indicates the linear size of the colloid. In the plot we set $r=0$ and all the other parameters to unity.}
    \label{fig:potentialtype}
\end{figure}

The prediction in Eq.~\eqref{eq:X2_first} is plotted in Figs.~\ref{fig:relax_A} and \ref{fig:relax_B}, which clearly show an initial exponential decay followed by a crossover towards the algebraic behavior, once the leading order contribution $\expval*{X^{(0)}(t)}=X_0 \exp(-\gamma t)$ has faded out. In Appendix \ref{par:generalpicture} we link the decay exponents for $r=0$ with the dynamical critical exponent $z=2+\alpha$ of the underlying Gaussian model; there we also derive the asymptotic form of the average position at long times
\begin{align}
    &\expval*{X_j (t)}  \simeq \frac{\sqrt{2\pi}\lambda^2}{e \nu k^2}  t^{-(d+2)/z} \times  \label{eq:mean_asymptotic} \\
    &\times \int \frac{\dd[d]{p}}{(2\pi)^d} p_j  |V_{pt^{-1/z}}|^2 \left(\vb{p} \cdot \vb{X_0}\right) \chi_{pt^{-1/z}}\left(t-1/\gamma\right) \; . \n
\end{align}
At criticality, $r=0$, this gives generically
\begin{align}
    \expval*{X(t)} 
 \simeq \frac{\lambda^2 c_1 X_0}{k}  \left(\gamma t\right)^{-1} \left(D t\right)^{-(d+2n)/z}  \sim t^{-1-(d+2n)/z} \; ,
    \label{eq:general_critical}
\end{align}
where $c_1$ is a numerical constant (see Eq.~\eqref{eq:asymptotic_critical} in Appendix \ref{par:generalpicture}), and the even integer $n$ indicates a coupling to the $n$-th derivative of the field, as in Eq.~\eqref{eq:derivativeinteraction}. For $n=0$, we recover from Eq.~\eqref{eq:general_critical} the critical exponents in Eqs.~\eqref{eq:relax_d_modelA} and \eqref{eq:relax_d} by setting $z=2$ (model A) or $z=4$ (model B), respectively. Specializing Eq.~\eqref{eq:mean_asymptotic} to the off-critical case of model B renders, instead,
\begin{align}
    \expval*{X (t)}  \simeq \frac{\lambda^2 c_2 X_0  D}{k\gamma} \left( D r \right)^{-(2+n+d/2)}  t^{-2-(d+n)/2}  \; ,
    \label{eq:general_massive}
\end{align}
where the numerical constant $c_2$ is given in Eq.~\eqref{eq:asymptotic_massive}. This dependence is, as expected, generically algebraic with a temporal decay to zero which is faster than the critical case. Figure~\ref{fig:couplingtype} shows how the value of $n$ changes the decay exponent of the asymptotic behavior in agreement with Eq.~\eqref{eq:general_critical}. Moreover, Eq.~\eqref{eq:mean_asymptotic} reveals that the details of the interaction potential $V_q$ do not affect the large-$t$ behavior of the tracer particle: indeed, the interaction potential only enters Eq.~\eqref{eq:mean_asymptotic} via $V_{pt^{-1/z}}\simeq V_{p\simeq 0}$, meaning that two distinct potentials with the same behavior for $p\simeq 0$ yield exactly the same asymptotic expression for the average position. This is verified in Fig.~\ref{fig:potentialtype}, where the average position is plotted for various choices of $V(\vb{x})$ and the corresponding curves become indistinguishable at long times.

In Section \ref{par:relevance} we will comment on how to amplify the long-time algebraic decay (which is most relevant at $r=0$) in possible experimental realizations of the system.

\begin{figure}
    \centering
    \includegraphics[]{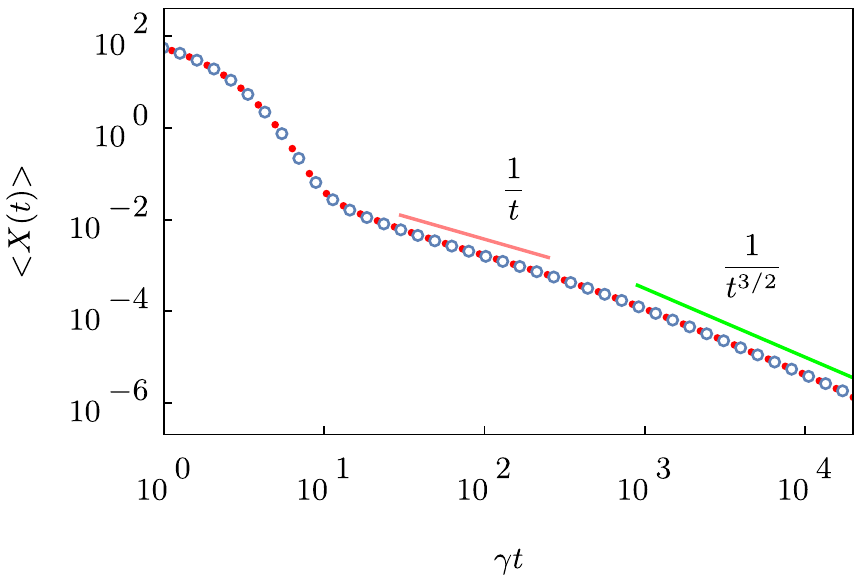}
    \caption{Relaxation of the average position $\expval*{X(t)}$ towards equilibrium in $d=1$ critical model A when the initial position $X_0$ is chosen sufficiently large so as to emphasize the nonlinear response. Open blue circles represent the theoretical prediction in Eq.~\eqref{eq:X2_first}, while red filled circles are the results of numerical simulations performed at $T=0$ (we justify this choice and describe the simulation method in Section \ref{par:numerical} and Appendix \ref{par:appendix_numerical}). Parameters used in the simulation are $\nu=1$, $k=0.1$, $X_0=150$, $D=1$, $R=1$, $\lambda=0.25$, integration timestep $\Delta t =0.01$ and lattice size $L=2048$.}
    \label{fig:nonlinear}
\end{figure}

One may ask the extent to which the results we obtained via a weak-coupling expansion could be retrieved by using a simpler linear response analysis. The linear response is formally recovered from Eq.~\eqref{eq:X2_first} as
\begin{equation}
    \expval*{X_j(t)}_\T{LR} \equiv X_0  \eval{\dv{X_0} \expval*{X_j\t(t)}}_{X_0=0}
\end{equation}
(the zeroth-order term trivially vanishes) and it turns out, with hindsight, that the long-time asymptotic expression in Eq.~\eqref{eq:mean_asymptotic} for the average position is indeed linear in $\vb{X}_0$. At short and intermediate times, however, nonlinear contributions arise which are encoded in the full response in Eq.~\eqref{eq:X2_first}, but they would be missed if we truncate it to the linear order. A simple way to highlight them is to choose $\vb{X}_0$ large enough so as to leave the linear response regime: Fig.~\ref{fig:nonlinear} shows the emergence of an intermediate algebraic behavior with different decay exponents, which is correctly described by Eq.~\eqref{eq:X2_first} and is actually observed in numerical simulations presented further below in Section \ref{par:numerical}. We give a semi-phenomenological description of this transient behavior in Section \ref{par:transient} and in Appendix \ref{app:nonlinear}; our analysis allows us to predict the amplitude and the slope of the average position in this regime, as well as an estimate of the crossover time $t_c$ at which the decay exponents in Eqs.~\eqref{eq:relax_d_modelA} and \eqref{eq:relax_d} are recovered.

\subsection{Comparison with the auto-correlation function}
The predictions presented above are to be compared with the long-time behavior of the auto-correlation function $C(t)\equiv\expval*{\vb{X}(t)\cdot \vb{X}(0)}$. The case of model B dynamics is discussed in Ref.~\cite{wellGauss}, where it was shown that
\begin{equation}
    \expval*{X(t)\cdot X(0)} \sim
    \begin{cases}
        t^{-d/4} &  \T{for}\;\; r=0 \; , \\
        t^{-\left(1+d/2 \right)} &  \T{for}\;\; r>0 \; .
    \end{cases}
    \label{eq:corr_d}
\end{equation}
 We briefly derive this result in Appendix \ref{par:autocorrelation} within the same perturbative framework as we did for the average position. In doing so, we extend the calculation to model A, for which we find
\begin{align}
    \expval*{X(t)\cdot X(0)} \sim
    \begin{cases}
        e^{-\gamma t}    &  \T{for}\;\; r>\gamma/D \; ,\\
        e^{-Drt}   &  \T{for}\;\; r<\gamma/D \; , \\
        t^{-d/2}   &  \T{for}\;\;  r=0 \; .
    \end{cases}
    \label{eq:corr_d_modelA}
\end{align}
The similarities between the two sets of exponents (see Eqs.~\eqref{eq:relax_d_modelA} and \eqref{eq:relax_d}) appear to be a manifestation of the fluctuation-dissipation theorem at long times. Indeed, one could write for the particle a linearized effective equation \cite{wellGauss} in the form
\begin{equation}
    \dot{X}(t) = F[X] + h(t) + \xi(t) \; ,
    \label{eq:effective}
\end{equation}
where $\xi(t)$ is white Gaussian noise, $h(t)$ is an external forcing term, and $F[X]$, possibly nonlocal in time, already contains the effects of the interaction with the field. The knowledge of the response function $R(\tau)$ would allow one to express, within the linear response regime,
\begin{equation}
    \expval*{X(t)} = \int_{t_0}^t \dd{t'} R(t-t')h(t') \; .
\end{equation}
Now, studying the relaxation of $X(t)$ starting from an initial condition $X_0\neq 0$ is tantamount to setting $h(t) = X_0 \delta(t-t_0)$ into the effective equation \eqref{eq:effective}, thus one concludes that
\begin{equation}
    \expval*{X(t)} = X_0 R(t) \; .
\end{equation}
Then it is clear that at long times, \ie, sufficiently close to equilibrium, the fluctuation-dissipation theorem holds, relating the linear response in Eqs.~\eqref{eq:relax_d_modelA} and \eqref{eq:relax_d} with the correlation function $C(t)$ in Eqs.~\eqref{eq:corr_d_modelA} and \eqref{eq:corr_d} according to
\begin{equation}
    R(t>0) = -\frac{1}{k_BT}\dv{C(t)}{t} \; .
\end{equation}

\section{Adiabatic approximation}
\label{par:Adiabatic}
In this Section we carry out a first-order adiabatic elimination of the field degrees of freedom which are assumed to be fast compared to the motion of the colloid: this way we obtain an effective equation for the dynamics of the particle alone. Note that projecting the fast degrees of freedom over the dynamics of the tracer particle adopting the Mori-Zwanzig scheme \cite{zwanzig,moriZwanzig}, which renders a linear equation, may lead to uncontrolled results in the present case, because the effective particle dynamics is actually nonlinear \cite{IOPmorizwanzig,jung2021nonmarkovian}. We follow instead Ref.~\cite{kaneko} and we \textit{integrate out} the field degrees of freedom using a transparent and physically intuitive procedure. In the process, we generalize the approach of Refs.~\cite{kaneko,theiss1,theiss2} to the case in which a continuum of fast variables are coupled to a single slow variable (see Appendix \ref{par:adiabatic} for further details). 

As it is customary in this context \cite{kaneko}, we will initially choose as a small parameter for the adiabatic expansion the ratio $\nu/D$ of the mobility of the particle to that of the field. However, it is clear from the discussion in Section \ref{par:Model} that the true time scale for the relaxation of the field variables is expressed by Eq.~\eqref{eq:tau_phi}, so that the long-wavelength Fourier modes exhibit slow relaxation close to criticality (model A) or even far from criticality for a conserved dynamics (model B). We thus expect the adiabatic approximation to eventually break down; in the following, we will be interested in locating when this breakdown occurs and possibly matching the adiabatic approximation with the weak-coupling solution in Eq.~\eqref{eq:X2_first}.

\subsection{Effective Fokker-Planck equation}
\label{par:effective_FP}
Let us go back to the coupled equations of motion \eqref{eq:particle} and \eqref{eq:fieldFourier} for the particle and the field, respectively. The first observation is that the equations for the Fourier components $\phi_q(t)$ decouple over the modes $q$: this holds true because we are considering the Gaussian model, which renders linear equations of motion. One should however bear in mind that the field $\phi(\vb{x},t)$ is real, which implies $\phi_q^*=\phi_{-q}$; this suggests to separate its real and imaginary parts $\phi_q^R \equiv \Re{\phi_q}$ and $\phi_q^I \equiv \Im{\phi_q}$ \cite{bettencourt}. We then rewrite Eqs.~\eqref{eq:particle} and \eqref{eq:fieldFourier} as
\begin{align}
    &\dot{\vb{X}}=-\gamma \vb{X} + \nu\lambda \int \dslash{q} \vb{q} \left( \phi_q^R g_q^I -  \phi_q^I g_q^R \right)  + \bm{\xi}(t) \; , \label{eq:coll_adiabatic} \\
    &\dot{\phi}_q^{R,I} = -\alpha_q \phi_q^{R,I} +D\lambda q^\alpha g_q^{R,I}  + \zeta_q^{R,I} \; , \label{eq:decoupled}
\end{align}
where we defined $g_q(\vb{X}) \equiv V_q \exp(-i\vb{q}\cdot \vb{X})$ and the noise correlations read
\begin{align}
    &\expval*{\zeta_q^{R,I}(t)\zeta_{q'}^{R,I}(t')}=  \frac{\Gamma_\phi}{2} \left[\delta^d(q-q') \pm \delta^d(q+q')\right]\delta(t-t') \; , \nonumber\\
    &\expval*{\zeta_q^R(t)\zeta_{q'}^I(t')} = 0 \; , \label{eq:noiseparts}
\end{align}
with $\Gamma_\phi \equiv 2DTq^\alpha$. The equations of motion for $\phi_q^{R,I}$ are now completely decoupled and their time-dependent probability distribution factorizes into
\begin{equation}
    \cor{P}\left[\phi,\vb{X},t\right] = \prod_{q \in \mathbb{R}^d} P\left( \phi_q^R;\vb{X},t \right) P\left( \phi_q^I;\vb{X},t \right) \; .
\end{equation}
Clearly, this $\cor{P}$ does not factorize into an $\vb{X}$-dependent and a $\phi$-dependent part, if not possibly at the initial time $t_0$. Note that the noise term in Eq.~\eqref{eq:noiseparts} still correlates $\phi_q^\sigma$ with $\phi_{-q}^\sigma$, for $\sigma=R,I$. When we write the Fokker-Planck equation corresponding to the set of Langevin equations \eqref{eq:coll_adiabatic} and \eqref{eq:decoupled}, this produces mixed derivatives in the form $\delta^2/(\delta \phi_{q}^\sigma \delta \phi_{-q}^\sigma)$, which can nonetheless be dealt with by noticing that $\phi_{-q}^R =  \phi_{q}^R$ and $\phi_{-q}^I = -\phi_{q}^I$. We thus obtain
\begin{equation}
    \partial_t \cor{P} = \left[ \cor{L}_X + \int \dslash{q} \left( \cor{L}_q^R + \cor{L}_q^I \right) \right] \cor{P} \; ,
    \label{eq:fpadiabatic}
\end{equation}
where, calling $\grad\equiv\grad_X$ and $\Gamma_x\equiv 2\nu T$, we introduced the operators
\begin{equation}
    \cor{L}_X = \div \left[ \gamma \vb{X} - \nu \lambda \int \dslash{q} \vb{q} \left( \phi_q^R g_q^I -  \phi_q^I g_q^R \right) \right] + \frac{\Gamma_x}{2} \laplacian
\end{equation}
and
\begin{eqnarray} 
    \cor{L}_q^\sigma = \fdv{\phi_q^\sigma} \left[ \alpha_q \phi_q^\sigma -D \lambda q^\alpha g_q^\sigma(\vb{X}) \right] + \frac{\Gamma_\phi}{2}\frac{\delta^2}{\delta (\phi_{q}^\sigma)^2} \; .
\end{eqnarray}
The second consideration is that the coupling with the field in the equation of motion for $\vb{X}$ is linear. The problem of the adiabatic elimination of a fast variable from a system of two stochastic differential equations was addressed, \eg, in Ref.~\cite{kaneko} and generalized in Refs.~\cite{theiss1,theiss2} to the case of a multi-dimensional Fokker-Planck equation linear in the fast variables. We sketch in Appendix \ref{par:adiabatic} how the same method can be naturally extended to Eq.~\eqref{eq:fpadiabatic}, which contains a continuum of fast variables. The resulting Fokker-Planck equation for the slow variable $\vb{X}(t)$ turns out to be
\begin{equation}
    \partial_t P(\vb{X},t) = \cor{L}_X^{\T{eff}}  P(\vb{X},t) \; ,
    \label{eq:finaladiabatic}
\end{equation}
where, in the case of an isotropic interaction potential,
\begin{equation}
    \cor{L}^{\T{eff}}_X = \div \left( \chi \gamma \vb{X} \right) +  \chi \nu T \laplacian + \order{\left(\frac{\nu}{D}\right)^2} \; ,
\end{equation}
with $\chi \equiv 1-\lambda^2 \mu$ and
\begin{equation}
    \mu \equiv \frac{\nu}{Dd} \int_\mathbb{R} \dslash{q} \frac{q^{2-\alpha}}{ (q^2+r)^2}|V_q|^2 \; .
    \label{eq:mu}
\end{equation}
This integral converges, at finite values of $r$, provided that the interaction potential $V_q$ decays sufficiently fast for large $q$, thus providing some form of ultra-violet cutoff. We may identify, \textit{a posteriori}, the coefficient $\lambda^2 \mu$ as the actual dimensionless small parameter in the adiabatic expansion which emerges naturally from the calculation.

Equation \eqref{eq:finaladiabatic} is markedly Markovian; non-Markovian effects would appear at the next perturbative order, here neglected \cite{kaneko}. It shows that, up to the second order in the adiabatic approximation, the only effect of the interaction with the field is to renormalize the drift and diffusion coefficients by the same amount in the equation of motion for an otherwise diffusing particle in a potential: this, in turn, is equivalent to rescaling time according to $t \rightarrow \chi t$. This is expected in order for Eq.~\eqref{eq:finaladiabatic} to render the correct steady state distribution $\cor{P}_\T{eq}\left(\vb{X}\right) \propto \exp(-\beta k X^2/2)$ of the particle, which does not depend on $\lambda$ (Appendix \ref{par:eqcolloid}). Such a dependence emerges instead during relaxation: in fact, Eq.~\eqref{eq:finaladiabatic} implies straightforwardly that a particle initially displaced from its equilibrium position at time $t_0=0$ will relax back as
\begin{equation}    
    \expval*{X_\T{ad}(t)} = X_0 e^{-\chi \gamma t} \; .
    \label{eq:x(t)adiabatic}
\end{equation}

\subsection{Comparison with the perturbative solution}
\label{par:match_main}
It is natural at this point to investigate if and when the perturbative solution in Eq.~\eqref{eq:X2_first} matches with the adiabatic approximation in Eq.~\eqref{eq:x(t)adiabatic}. In order to address this issue, we consider their ratio
\begin{equation}
    \zeta \equiv \frac{ \expval*{X_\T{ad}(t)}}{\expval*{X(t)}}-1 = \lambda^2 \left[ \mu \gamma t - \frac{\expval*{X\t(t)}}{X_0}e^{\gamma t} \right] +\order{\lambda^4} \; ,
    \label{eq:balancing}
\end{equation}
which vanishes when the adiabatic approximation gives the same result as the weak-coupling expression at this perturbative order. Note that $\zeta$ can be computed analytically by choosing, for instance, a Gaussian or $\delta$-like potential, as done in Appendix \ref{par:matching}. In order for $\zeta$ to vanish for some time $t$, and therefore for the adiabatic approximation to be accurate, we need $\expval*{X\t(t)}e^{\gamma t}$ to be linear in $t$. One might expect this to be the case at long times $t$: indeed, the colloid moves faster initially, when it is released, while it slows down as it reaches the bottom of the harmonic trap, thus making heuristically the adiabatic approximation more reliable. We have already analyzed the behavior of $\expval*{X\t(t)}$ in this regime both for model A in Eq.~\eqref{eq:relax_d_modelA}, and model B in Eq.~\eqref{eq:relax_d}, so we conclude that:
\begin{enumerate}[(i)]
    \item{The adiabatic approximation is never accurate in model B: indeed, $\expval*{X\t(t)}$ always decays algebraically at large $t$ and there is no way that it can counterbalance the term $e^{\gamma t}$, thus causing $|\zeta|$ to grow without bounds. This is not surprising, because in the whole adiabatic elimination procedure we have used the ratio of the two mobilities $\nu/D$ as a small adiabaticity parameter; however, the actual timescale $\tau_\phi$ for the relaxation of the field is given in Eq.~\eqref{eq:tau_phi}, which shows that, for any choice of $D$ and $r$, there are always long-wavelength Fourier modes in model B which relax slower than the colloid.}
    \item{By the same token, the timescale for relaxation in model A is given by Eq.~\eqref{eq:tau_phi} with $\alpha=0$, so that the slowest mode is characterized by $\tau_\phi^{-1}(q=0) = Dr$. We are led to the conclusion that $\expval*{X\t(t)}e^{\gamma t}$ can only possibly behave linearly when $Dr>\gamma$, as it is clear by looking at Eq.~\eqref{eq:relax_d_modelA}. Being $\tau_X^{-1} = \gamma$ the timescale of relaxation of the colloid in the trap, this implies that even the slowest field mode must relax faster than the colloid.}
\end{enumerate}
In Appendix \ref{par:matching} we determine, in the case of model A, the linear growth coefficient $a$ defined as
\begin{equation}
    \expval*{X\t(t)}e^{\gamma t} \simeq a t \;\;\;\;\;\; \T{for}\;\;t\gg \tau_X \; ,
\end{equation}
which enters the definition of $\zeta$ in Eq.~\eqref{eq:balancing}; we then compare it to the values of $\mu$ in Eq.~\eqref{eq:mu} computed with the same interaction potential (which is chosen to be Gaussian for definiteness). This way we prove that the balancing in Eq.~\eqref{eq:balancing} does occur, thus making $\zeta=0$ at long times.

\begin{figure*}
\centering
\subfloat{
  \centering
  \includegraphics[]{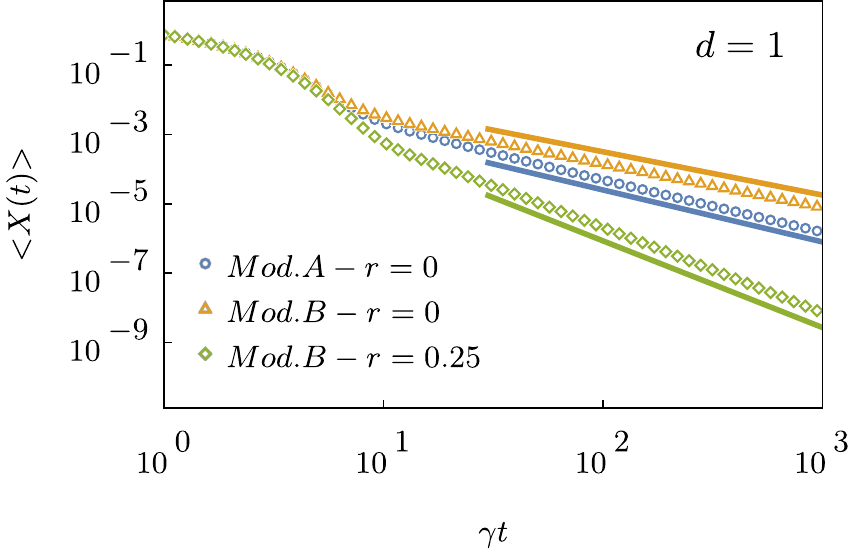}
}
\subfloat{
  \centering
  \includegraphics[]{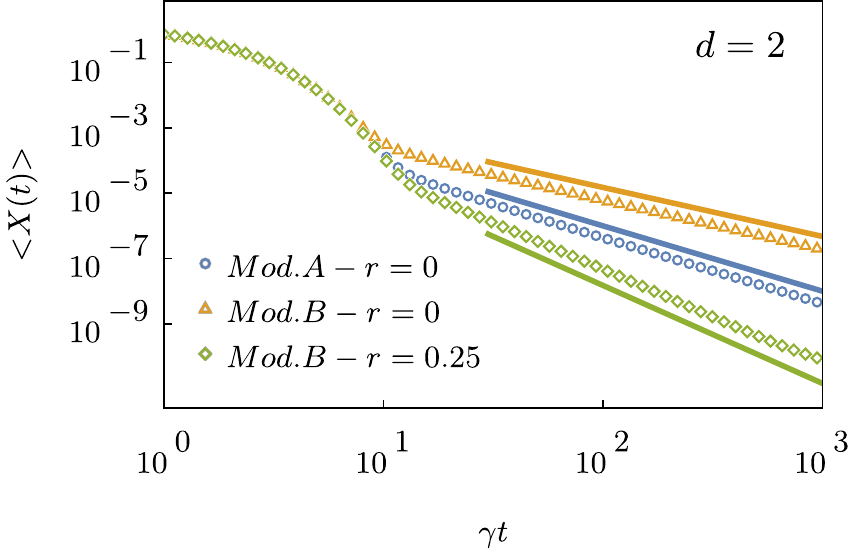}
  }
\caption{Average position $\expval*{X(t)}$ of the particle in numerical simulations of the noiseless equations of motion, corresponding to $T=0$, in $d=1$ (left) and $d=2$ (right). All the simulations are in excellent agreement with the long-time behavior predicted in Eqs.~\eqref{eq:relax_d_modelA} and \eqref{eq:relax_d} (corresponding to the slopes indicated by the solid straight lines). We do not show here the full prediction in Eq.~\eqref{eq:X2_first} for graphical clarity, as it is almost indistinguishable from the simulation points (but we do present such a comparison in Figs.~\ref{fig:nonlinear} and \ref{fig:fits}). The parameters used in the simulation are $\nu=1$, $k=0.1$, $X_0=2$, $D=1$, $R=0.5$, $\lambda=0.25$, and $\Delta t =0.01$. The system size is chosen to be $L=2048$ in the $d=1$ case, and $L=512$ in the $d=2$ case.}
\label{fig:noiseless}
\end{figure*}

\begin{figure}
    \centering
    \resizebox{\linewidth}{!}{\includegraphics[]{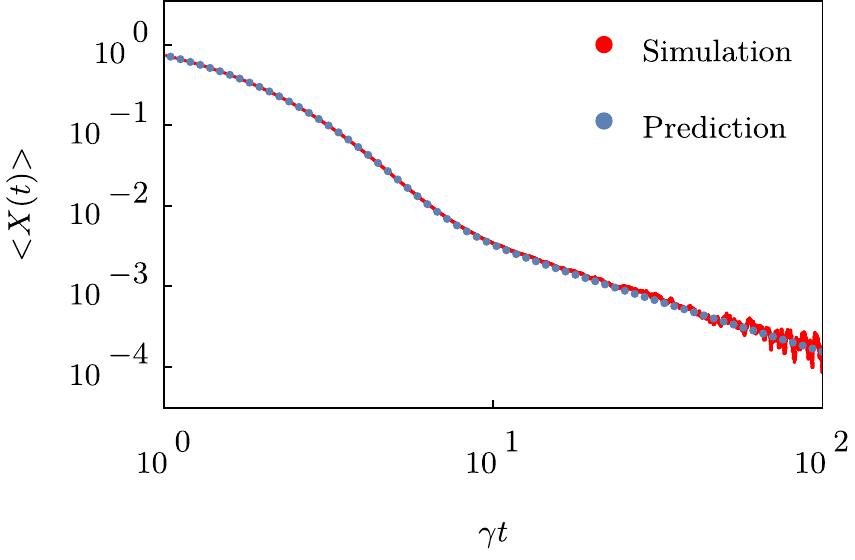}}
    \caption{Average particle position $\expval*{X(t)}$ during the relaxation to equilibrium in $d=1$ critical model B, in the presence of noise. Simulation results are plotted as a solid red line, while the blue dots represent the theoretical prediction in Eq.~\eqref{eq:X2_first}; they are shown to be in complete agreement. Parameters used in the simulation are $r=0$, $\nu=1$, $k=0.1$, $X_0=2$, $D=1$, $R=1$, $\lambda=0.25$, $\Delta t =0.01$, $T=0.1$, $L=128$, and $N=7.7\times 10^{8}$ realizations.}
    \label{fig:fits}
\end{figure}

\section{Numerical simulation}
\label{par:numerical}
In order to verify the validity of our analytical predictions beyond the various approximations considered, we numerically simulate the system by direct integration of the coupled Langevin equations of motion for the field and the particle, Eqs.~\eqref{eq:field} and \eqref{eq:particle}, respectively. To this end, we discretize the field over a lattice of size $L$ and we adopt periodic boundary conditions, as described in Appendix \ref{par:appendix_numerical}. A great simplification arises by noticing that the long-time asymptotic expression we found in  Eq.~\eqref{eq:mean_asymptotic} for the average position of the colloid does not depend on the temperature $T$ (which affects instead the dynamics at intermediate times and the amplitude of the thermal fluctuations). At long times and close to the equilibrium position $\vb{X}=0$, noise fluctuations make it challenging to observe clearly the algebraic decay predicted in Eqs.~\eqref{eq:relax_d_modelA} and \eqref{eq:relax_d}. In addition, it is well-known that very large systems are needed in order to sample the vicinity of a bulk critical point without incurring in finite-size effects. Accordingly, we first simulate the noiseless equations of motion, corresponding to setting $T=0$, in large systems in $d=1$ and $d=2$, finding excellent agreement with the analytical prediction in Eq.~\eqref{eq:X2_first} and its long-time algebraic behavior. This is presented in Fig.~\ref{fig:noiseless}, which shows the average position in simulations performed at small values of the coupling $\lambda$ (solid lines represent the slope of the long-time algebraic behavior predicted by  Eqs.~\eqref{eq:relax_d_modelA} and \eqref{eq:relax_d}). We then focus on one of these curves and we re-introduce the noise by considering $T\neq 0$, showing that in fact the effect of thermal fluctuations on the average colloid displacement is negligible provided that one averages over a sufficiently large number $N$ of realizations. Indeed, we show in Fig.~\ref{fig:fits} that even the \textit{noisy} curve agrees with the prediction in Eq.~\eqref{eq:X2_first}, with scarce dependence on the specific choice of the interaction potential $V_q$, provided that its characterizing length scale $R$ is of the same order as the one used in the simulation (which is performed by adopting a Gaussian interaction potential, see Appendix \ref{par:appendix_numerical}).

\subsection{Analysis of the transient behavior for large $X_0$}
\label{par:transient}
\begin{figure}
    \centering
    \resizebox{\linewidth}{!}{\includegraphics[]{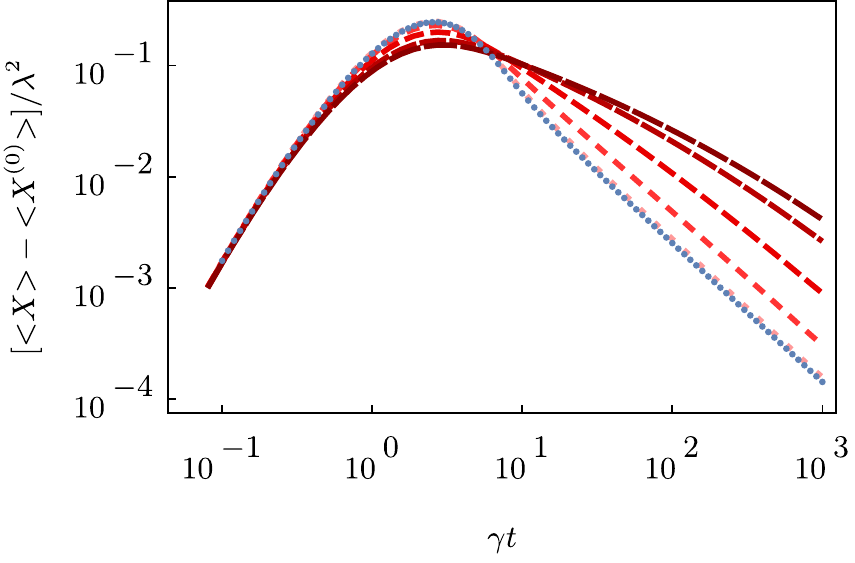}}
    \caption{Perturbative correction to the average position $\expval*{X(t)}$ during the relaxation to equilibrium in $d=1$ critical model B. Blue dots represent the theoretical prediction of $\expval*{X^{(2)}(t)}$ in Eq.~\eqref{eq:X2_first}, while we plotted in different shades of red (and different dashing) the quantity $[\expval*{X}-\expval*{X^{(0)}}]/\lambda^2$ estimated in numerical simulations for increasing values of $\lambda\in [0.25-2.00]$, from lightest to darkest (and from shortest to longest dashing). For each curve we subtracted from the data the purely exponential decay and divided by $\lambda^2$. For large values of $\lambda$, one observes qualitatively the same power-law decay at long times, whose onset is nonetheless delayed as $\lambda$ increases. Parameters used in the simulation are $T=0$, $\nu=1$, $k=0.1$, $X_0=2$, $D=1$, $R=1$, $\Delta t =0.01$, and $L=128$.}
    \label{fig:biglambda}
\end{figure}

As anticipated in Section \ref{par:longtime}, by choosing a sufficiently large value of the initial displacement $X_0$ one observes an intermediate, algebraic behavior in the average particle position, highlighted in Fig.~\ref{fig:nonlinear}. This would not be captured by a linear response analysis of the system, but it is correctly described by the perturbative prediction in Eq.~\eqref{eq:X2_first}. In this Section we use such analytical prediction together with numerical simulations of the system in order to provide a phenomenological description of this transient behavior within the small-$\lambda$ regime, where Eq.~\eqref{eq:X2_first} agrees well with numerical data. By inspecting several relaxation curves corresponding to different values of the initial displacement $X_0$, one can observe the following:
\begin{enumerate}[(i)]
    \item For short times $t\ll \tau_X=\gamma^{-1}$, the dynamics is dominated by the initial exponential decay determined by the force exerted by the harmonic trap. If one insists on isolating the $\order{\lambda^2}$ correction to the average position by subtracting the leading order exponential decay, they would observe an initial growth (qualitatively analogous to Fig.~\ref{fig:biglambda}) whose precise form is influenced by all the microscopic details of the confining potential and of the interaction potential, such as $\gamma$, $R$ and the functional form of $V(\vb{x})$ (see, \eg, Fig.~\ref{fig:potentialtype}).
    \item For $t \gtrsim \tau_X$ and up to a crossover time which we denote by $t_c$, the average displacement of the colloid decays algebraically with an exponent which does not coincide with the one eventually displayed at longer times. This exponent shows some universal features, as it only depends on the spatial dimensionality of the system and on the critical properties of the field (\ie, on its dynamical critical exponent $z$). Moreover, quite surprisingly, the amplitude of $\expval*{X(t)}$ in this regime turns out to be independent of the value of $X_0$ itself, a clear example of nonlinear response.
    \item For $t > t_c$, we recover the asymptotic decay exponents predicted by Eqs.~\eqref{eq:relax_d_modelA} and \eqref{eq:relax_d}, in agreement with linear response analysis. The crossover time $t_c$ becomes larger upon increasing $X_0$. 
\end{enumerate}
The problem is analyzed in full details in Appendix \ref{app:nonlinear}. We start by identifying the crossover time $t_c$ with the relaxation timescale of the field over length scales comparable with $X_0$: this timescale can be read in Eq.~\eqref{eq:tau_phi} by setting $q\sim 1/X_0$, which yields in the critical case $t_c \sim X_0^z/D$. The physical motivation is the following. At time $t=0$ the colloid is released in position $X_0$ and enters in contact with the field; since the latter has a nonzero relaxation time, at short times $t\lesssim \tau_X$ we expect the particle to be dragged primarily by the restoring force of the harmonic trap, $\dot{X}\simeq -\gamma X_0$. On a timescale given by $\tau_X=\gamma^{-1}$ the colloid covers a distance of the order of $\Delta X \sim X_0$, so that it becomes relevant to consider the time $t_c(X_0)$ taken by the field in order to rearrange over such a distance. Once the field has reached a state close to its equilibrium configuration around the colloid (which is by now close to the center of the harmonic trap), then the dynamics is captured by linear response and we recover the asymptotic results of Section \ref{par:longtime}. Of course the transient regime cannot be appreciated if one chooses a small value of $X_0$, simply because correspondingly $t_c \ll \tau_X$.

Motivated by the phenomenological observation stated above that the behavior of the particle is algebraic within the transient region $t\lesssim t_c$, while the amplitude is independent of $X_0$, we propose for times $t\gg \tau_X$ the scaling ansatz
\begin{equation}
     \expval*{X(t)} \simeq c_0 t^{-\alpha_0} f\left( t/t_c \right) \; ,
     \label{eq:scaling_transient}
\end{equation}
where $f(\tau)$ is a scaling function with the property that
\begin{equation}
    f(\tau) \sim
    \begin{cases}
        \tau^{-\beta_0} & \T{for}\;\;  \tau\gg 1 \; , \\
        \T{const.} & \T{for}\;\; \tau\lesssim 1 \; .
    \end{cases}
\end{equation}
The intermediate exponent $\alpha_0$ and the coefficient $c_0$ (the latter up to some numerical constant) can now be determined from an asymptotic matching of Eq.~\eqref{eq:scaling_transient} with the long-time expression $\expval*{X_j (t)} \simeq c_\infty X_0 \, t^{-\alpha_\infty}$, where $c_\infty$ and $\alpha_\infty$ are known from our previous asymptotic calculation, Eqs.~\eqref{eq:mean_asymptotic} and \eqref{eq:general_critical}. This gives at criticality
\begin{equation}
    \alpha_0 = 1+\frac{d-1}{z} \;, \;\;\;\;\; \T{and} \;\;\;\;\; c_0 \propto \frac{\lambda^2}{\gamma k}D^{(1-d)/z} \; .
\end{equation}
A similar analysis can be repeated for the off-critical case in model B, yielding for $\xi \ll X_0$ a crossover time $t_c \sim X_0^2/(D r)$ with intermediate exponent and proportionality factor
\begin{equation}
    \alpha_0 = 2+\frac{d-1}{2} \;, \;\;\;\;\; \T{and} \;\;\;\;\; c_0 \propto \frac{\lambda^2 D}{\gamma k}\left(D r \right)^{-(d+3)/2} \; ,
\end{equation}
respectively.

\begin{figure}
    \centering
    \includegraphics[]{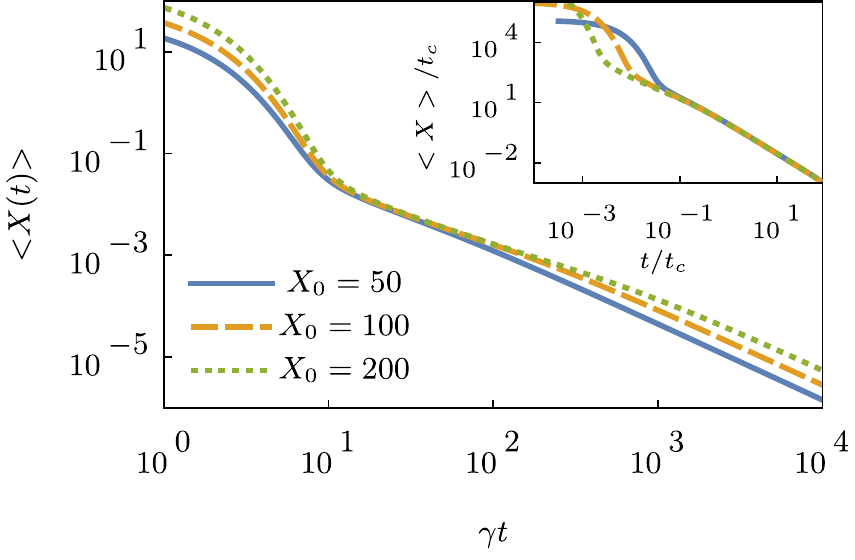}
    \caption{Average particle position $\expval*{X(t)}$ during its relaxation towards equilibrium in $d=1$ critical model A, when the initial position $X_0$ is chosen sufficiently large so as to emphasize the nonlinear response. In the main plot, the various curves correspond to increasing values of $X_0$, and the associated crossover time $t_c$ is seen to shift towards larger times. In the inset, the same curves are collapsed according to the scaling form in Eq.~\eqref{eq:scaling_transient} (see the main text). Parameters used in the simulation are $\nu=1$, $k=0.1$, $T=0$, $D=1$, $\lambda=0.25$, $R=1$, $\Delta t =0.01$, and $L=8192$.}
    \label{fig:transient}
\end{figure}

In Fig.~\ref{fig:transient} we plot the average position of the tracer particle in the case of critical model A ($d=1$) for three values of the initial displacement $X_0$. In the main plot we observe that the three curves share the same amplitude within the transient region $\tau_X \ll t \lesssim t_c$, with $t_c$ becoming larger as $X_0$ is increased. In the inset we exhibit the collapse of the three curves according to the scaling ansatz in Eq.~\eqref{eq:scaling_transient}, which can equivalently be written as $ \expval*{X_j(t)} \simeq c_0 t_c^{\alpha_0} f_2\left( \tau \right) $ upon defining $\tau=t/t_c$ and $f_2(\tau) \equiv \tau^{-\alpha_0} f(\tau)$. Plotting $t_c^{-\alpha_0} \expval*{X_j(t)} \,\T{vs}\, (t/t_c)$ shows indeed that a single curve $f_2\left( \tau \right)$ well describes the dynamics for $t\gg\tau_X$.

\subsection{How to amplify the long-time algebraic decay}
\label{par:relevance}
Here we address the question of how to control the overall amplitude of the algebraic decay predicted for the average particle position. Indeed, although our model is not meant to describe the dynamics of an actual colloid in a fluid, it still makes sense to check whether it would be in principle possible to amplify it and make it comparable with the length scale of the colloid radius $R$. A naive look at the asymptotic expressions in Eqs.~\eqref{eq:general_critical} and \eqref{eq:general_massive}, which are linear in $X_0$, would lead to the (wrong) conclusion that the algebraic decay can be enhanced by increasing $X_0$. However, we have checked in Section \ref{par:transient} that the crossover time $t_c$ at which the asymptotic algebraic decay starts to be seen increases upon increasing $X_0$. Accordingly, one should better ask: how large is the average position at time $t_c$, when the decay assumes its asymptotic algebraic form? Interestingly, plugging the various estimates for $t_c$ given in Section \ref{par:transient} into Eqs.~\eqref{eq:general_critical} and \eqref{eq:general_massive} leads to the same expression for the position at the crossover time, \ie,
\begin{align}
    \expval*{X (t_c)} = \frac{\lambda^2 c D}{k\gamma} X_0^{1-d-z},
    \label{eq:crossover_amplitude}
\end{align}
where the numerical constant $c$ is either $c_1$ or $c_2$ for the critical or off-critical cases, respectively. This expression tells us that the optimal value of $X_0$ should be chosen as small as possible in order to amplify the effect, but still sufficiently large so as to satisfy the assumption $t_c > \tau_X$ introduced in Section \ref{par:transient}.

We now recall that the coupling parameter $\lambda$ is not dimensionless, so that the notion of ``small $\lambda$'' we have often adopted in the previous Sections has to be made more precise. To do this, we now choose to measure lengths in units of the colloid radius $R$. The position at time $t_c$ can then be conveniently expressed as
\begin{align}
    \expval*{X (t_c)} = g^2 R \left( X_0/R \right)^{1-d-z},
    \label{eq:crossover_amplitude_dimensionless}
\end{align}
where the dimensionless coupling $g^2\equiv c \lambda^2 D/(k \gamma R^{d+z})$ emerges naturally as the actual small parameter for our perturbative expansion in Eq.~\eqref{eq:Xseries}. \footnote{An upper bound on the value of $\lambda$ can in principle be obtained by requiring $\expval*{X (t_c)}\ll X_0$ from Eq.~\eqref{eq:crossover_amplitude}. However, this bound is generally too loose to be of practical use.}

In Appendix \ref{app:relevance} we focus on the case of the off-critical model B, which is the closest to experimental realizations among the models we considered in this work. Choosing for the various parameters of the model the typical values corresponding to experiments with silica particles immersed in binary fluid mixtures \cite{GambassiCCF,energytransfer}, we show that the amplitude of the effect we predicted in Eq.~\eqref{eq:crossover_amplitude} is in principle well within the reach of digital videomicroscopy.

\begin{figure}
    \centering
    \resizebox{\linewidth}{!}{\includegraphics[]{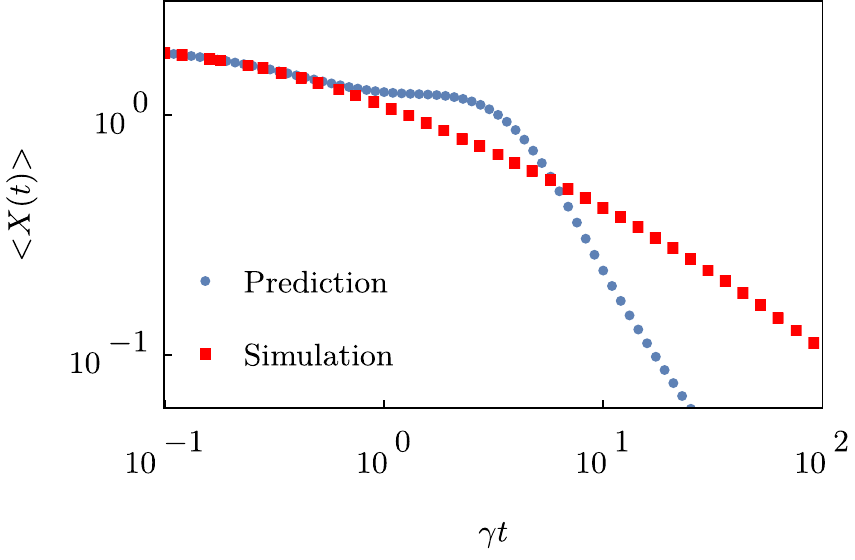}}
    \caption{Average particle position $\expval*{X(t)}$ during its relaxation to equilibrium in $d=1$ critical model B. We chose a large value of the coupling constant $\lambda$, well beyond the perturbative regime where agreement is observed between simulation data and our analytical prediction. Here the theoretical prediction indeed fails to describe even the qualitative behavior of the average position at short times (see main text). Parameters used in the simulation are $\lambda=2$, $T=0$, $\nu=1$, $k=0.1$, $X_0=2$, $D=1$, $R=1$, $\Delta t =0.01$, and $L=128$.}
    \label{fig:biglambda_comparison}
\end{figure}

\subsection{A hint at the large-$\lambda$ behavior}
\label{par:large_lambda}
The agreement between the perturbative solution in Eq.~\eqref{eq:X2_first} and the numerical simulations justifies the weak-coupling approximation we adopted throughout this work and ensures that the higher-order contributions which we have systematically neglected do not become increasingly relevant at long times, at least as long as the coupling constant $\lambda$ is small. Now we can use the numerical simulation to explore the regime in which $\lambda$ becomes larger. We will consider for definiteness the case of critical model B in $d=1$ and choose values of the coupling $\lambda\in [0.25-2.00]$. With the choice of parameters $k=0.1$ and $D$, $\gamma$, and $R$ set to unity, this corresponds to taking the dimensionless coupling $g$ defined in Section \ref{par:relevance} within the range $g \in [0.55-4.42]$.

Figure~\ref{fig:biglambda} compares the prediction in Eq.~\eqref{eq:X2_first} with the corresponding total correction to the average position, including higher-orders, which we can extract from the simulation data by subtracting from the measured trajectory $\expval{X(t)}$ the purely exponential decay $\expval{X\z(t)}$ predicted at $\order{\lambda^0}$, and therefore dividing by $\lambda^2$. One observes that at long times the exponent of the algebraic decay does not change upon increasing $\lambda$, but the amplitude predicted by Eq.~\eqref{eq:mean_asymptotic} acquires positive corrections coming from higher-order contributions. The time at which the onset of the power-law behavior occurs also shifts towards longer times as the value of $\lambda$ increases. 

A common feature in all the curves shown in Fig.~\ref{fig:biglambda} is that the $\order{\lambda^2}$ correction, which vanishes at $t=0$, grows up to a maximum value before decaying algebraically to zero. One can envision that, for large enough $\lambda$, the correction $\lambda^2 \expval*{X^{(2)}(t)}$ would become larger than the leading term $\expval*{X^{(0)}(t)}$, thus affecting its monotonic behavior. Of course such a scenario is well beyond the reach of the asymptotic expansion in Eq.~\eqref{eq:Xseries}, and in fact it is proven wrong in the numerical simulations performed at large $\lambda$ which we report in Fig.~\ref{fig:biglambda_comparison}, where a clear departure from the weak-coupling prediction is observed even at short times. 

In passing, we observe that the initial growth of the correction $\expval*{X^{(2)}(t)}$ to the average position shown in Fig.~\ref{fig:biglambda} also presents an algebraic behavior (although, of course, the effect is masked by the leading exponential contribution in this short-time regime). However, the characterizing exponents are found in this case to depend on the specific choice of the interaction potential $V(\vb{x})$, while they are in general insensitive to the value of $r$ quantifying the distance from criticality.

\section{Summary and conclusions}
\label{par:conclusion}
We analyzed the relaxation towards equilibrium of a colloidal particle linearly coupled to a scalar Gaussian field, both following a stochastic evolution which preserves detailed balance at all times. Working within a weak-coupling expansion, we have shown that the average position of the particle displays an algebraic decay at long times (see Eqs.~\eqref{eq:relax_d_modelA} and \eqref{eq:relax_d}) when the field is close to its bulk critical point (Figs.~\ref{fig:relax_A} and \ref{fig:relax_B}), and also far from criticality for a conserved-type field dynamics (model B). At criticality, we related these decay exponents with the dynamical critical exponent $z$ of the underlying Gaussian dynamical field theory, see Eq.~\eqref{eq:general_critical}. These exponents exhibit a certain degree of universality, in the sense that they depend only on the spatial dimensionality of the system but not on the specific form of the coupling between the field and the particle, provided that it is linear and translationally invariant. 
We supported these predictions beyond the perturbative approximation through numerical integration of the Langevin equations of motion, as shown in Figs.~\ref{fig:noiseless} and \ref{fig:fits}.

In the adiabatic limit, we derived an effective Fokker-Planck equation for the colloidal particle by integrating out the field degrees of freedom from the coupled equations of motion; then we used it in order to obtain an adiabatic approximation of the relaxation towards equilibrium in the same setting. The matching of the adiabatic solution with that obtained via weak-coupling approximation is only possible for a dissipative field dynamics (model A) and sufficiently far from criticality so that $\tau_\phi \ll \tau_X$, being $\tau_\phi$ and $\tau_X$ the relaxation timescales of the noninteracting field and of the particle, respectively (Eqs.~\eqref{eq:tau_x} and \eqref{eq:tau_phi}). In particular, since $\tau_\phi$ can become arbitrarily large at the critical point due to the presence of long-wavelength modes, the adiabatic approximation can \textit{never} be applied for a critical field in the bulk, as it was heuristically expected. In the case of a conserved field dynamics (model B), moreover, the adiabatic approximation fails also away from criticality, because of the presence of such slow modes for any value of the parameter $r$.

Finally we showed that, by choosing a sufficiently large value of the initial displacement $X_0$, a transient algebraic regime is observed in the average position of the particle which would be entirely missed if one had adopted linear response analysis, while it is correctly described by our perturbative prediction in Eq.~\eqref{eq:X2_first}. The main features of this intermediate regime are encoded in the scaling form we proposed in Eq.~\eqref{eq:scaling_transient}.

We emphasize that the conclusions we reached in this work are in principle qualitatively testable with current experimental technology, for instance by microscopic observation of silica particles trapped by optical tweezers and immersed in a binary liquid mixture close to the critical point of the demixing transition \cite{casimirColloids,energytransfer}.

Various related problems can be addressed within this model. A question we left open in this work is for instance whether the inclusion in the Hamiltonian of additional terms which are nonlinear in the field $\phi$ may have an effect on the decay exponents of the average particle position. We expect the latter to depend in general on the static and dynamic universality classes of the bulk field Hamiltonian, so that additional terms involving $\phi$ (and not $\vb{X}$) should play a role whenever they are relevant in the renormalization group sense. On the other hand, nonlinear couplings such as $\sim \phi^3(\vb{x})V(\vb{x}-\vb{X})$ (which have the same symmetry as those considered in this work) may turn out to provide subleading contributions at long times. The case of relaxation towards equilibrium is possibly the simplest nonequilibrium scenario, but another typical setting is that of an external periodic forcing which drives the particle into a nonequilibrium periodic state \cite{2parts}. Moreover, if two or more particles interact with the same fluctuating field, then an effective interaction between them arises which modifies both their static and dynamical behavior. Different types of field-particle couplings can also be studied, as well of the effects of the field on the dynamics of an active tracer particle \cite{PhysRevLett.121.028001,active}. Finally, in order to approach the critical point of the fluid we are trying to model, we are naturally led to go beyond the Gaussian approximation and use, instead, a scalar $\phi^4$ theory as a starting point for a perturbative analysis, as well as more appropriate minimal models of the dynamics. We shall explore these issues in future works.

\begin{acknowledgments}
We thank U. Basu for useful discussions. FF acknowledges the support received from the University of Trento and SISSA. DV would like to thank A. Galvani, G. Giachetti and L. Sesta for illuminating discussions, and B. Walter, who is also co-writer of the code used for numerical simulations and run on Ulysses SISSA computing facilities. AG acknowledges support from MIUR PRIN project “Coarse-grained description for non-equilibrium systems and transport phenomena (CO-NEST)” n. 201798CZL.
\end{acknowledgments}

\bibliography{references} 

\onecolumngrid
\appendix

\section{Correlation functions}
\label{par:correlators}
In this Appendix we calculate the expectation values over the decoupled processes in Eqs.~\eqref{eq:field} and \eqref{eq:particle}, which are recovered by setting $\lambda=0$.

\subsection{One and two-time correlation functions of the non-interacting particle}
Each component $X_j$ of the particle position $\vb{X}(t)$ is ruled at $\order{\lambda^0}$ by a Gaussian and Markovian process (see Eq.~\eqref{eq:particle}). Accordingly, its propagator is Gaussian:
\begin{equation}
    P(\vb{X},t|\vb{X}_0,t_0) = \left[ \frac{1}{\sqrt{2\pi}\sigma} \right]^d \exp[-\frac{|\vb{X}-\vb{m}(t)|^2}{2\sigma^2(t,t_0)}] \; ,
\end{equation}
where, denoting by $\expval*{A|B}$ the conditional average,
\begin{equation}
    \vb{m}(t) = \vb{m}(t;\vb{X}_0,t_0) \equiv \expval*{\vb{X}(t) | \vb{X}(t_0) = \vb{X}_0} \; ,
\end{equation}
and the variance is the same for each component $X_j$, due to the isotropy of the problem:
\begin{equation}
    \sigma^2(t,t_0) \equiv \expval*{X_j^2(t) | X_j(t_0) = (\vb{X_0})_j} - m_j^2(t) \; .
\end{equation}
At $\order{\lambda^0}$ the particle position is described by the Ornstein-Uhlenbeck process (\ie, by Brownian diffusion in a harmonic potential \cite{risken}), for which it is easy to derive the well-known results
\begin{align}
    \vb{m}(t) &= \expval*{\vb{X}^{(0)}(t)} =\vb{X}_0e^{-\gamma(t-t_0)} \; ,
    \label{eq:m(t)} \\[2mm]
     \sigma^2(t,t_0) &= \frac{T}{k}\left[ 1-e^{-2\gamma(t-t_0)} \right] \; ,
\end{align}
where we introduced $\gamma\equiv \nu k$ as in Section \ref{par:Model} and we assumed the particle to start its motion at position $\vb{X}^{(0)}(t_0)=\vb{X}_0$ at time $t=t_0$. Similarly, the connected two-time correlation function $C(t_1,t_2)$ is given by
\begin{align}
        C(t_1,t_2) &\equiv \expval*{X_j\z (t_1) X_j\z (t_2)}_c = \expval*{\left[ X_j\z (t_1) - \expval*{X_j\z (t_1)} \right]\left[ X_j\z (t_2) - \expval*{X_j\z (t_2)} \right] } \n \\
        & = \frac{T}{k} \left[ e^{-\gamma|t_2-t_1|} - e^{-\gamma(t_1+t_2-2t_0)} \right] \; .
    \label{eq:2point}
\end{align}

\subsection{Two-time correlation function of the non-interacting field}
\label{par:freefield}
The Langevin equation \eqref{eq:fieldFourier} for the field reads at $\order{\lambda^0}$ and in Fourier space
\begin{align}
    \dot{\phi}\z_q &= -\alpha_q \phi_q\z + \zeta_q \; , \\[2mm]
    \expval*{\zeta_q(t)\zeta_{q'}(t')} &= 2DTq^\alpha \delta^d(q+q')\delta(t-t') \; ,
\end{align}
where $\alpha_q= Dq^\alpha(q^2+r)$, while $\alpha=2$ for model B and $\alpha=0$ for model A.
This equation is formally identical to that of the Ornstein-Uhlenbeck particle, so it is easy to derive \cite{Janssen1989}
\begin{equation}
    \expval*{\phi_q\z(s_1)\phi_p\z(s_2)} = \delta^d(p+q)\left[ C_q^D(s_1,s_2) + G_q(s_1-t_0)G_q(s_2-t_0)\phi_q^2(t_0) \right] \; ,
    \label{eq:field_corr_general}
\end{equation}
where we introduced the free-field propagator $G_q(s)=\theta(s)e^{-\alpha_q s}$. Here $\theta(s)$ is the Heaviside step function, and
\begin{equation}
    C^D_q(s_1,s_2) = \frac{T}{q^2+r} \left[ e^{-\alpha_q |s_2-s_1|} - e^{-\alpha_q (s_1+s_2-2t_0)} \right]
    \label{eq:freefield}
\end{equation}
is the correlation function corresponding to the case of Dirichlet initial condition $\phi_q(t_0)\equiv0$. It also coincides with the connected correlation function $\expval*{\phi_q\z(s_1)\phi_p\z(s_2)}_c$ computed with any other \textit{fixed} initial condition $\phi_q(t_0)$. For $t_0\rightarrow -\infty$ we recover from Eq.~\eqref{eq:field_corr_general} the correlation function in the stationary state, which is time-translational invariant, as expected.

If we assume that $\phi_q(t_0)$ is randomly drawn from the stationary distribution of the field and we average the correlation function in Eq.~\eqref{eq:field_corr_general} over all possible initial conditions, we get $\expval*{\phi_q^2(t_0)}_\T{i.c.}=T/(q^2+r)$ and it follows that
\begin{equation}
\expval*{\phi_q\z(s_1)\phi_p\z(s_2)}_\T{i.c.} = \delta^d(p+q) C_q(s_2-s_1) \; ,
\end{equation}
where
\begin{equation}
    C_q(\tau) = \frac{T}{q^2+r} e^{-\alpha_q\abs{\tau}}
    \label{eq:field_corr}
\end{equation}
is the expected equilibrium correlation function of the field.
This is the expression we will adopt in this work, as we assume the field to be initially in thermal equilibrium, before the particle is added.

 \subsection{$n$-time correlation functions of the non-interacting particle}
\label{par:expaverages}
The knowledge of the one- and two-time correlation functions discussed above is sufficient in order to write down the generating functional $\cor{Z}[j]$ of the $n$-time correlation functions for the Ornstein-Uhlenbeck (or any other Gaussian) process: for each spatial component $x=x_i$ separately, it reads
\begin{align}
        \cor{Z}[j] &= \expval*{e^{\int_{-\infty}^\infty \dd{s} j(s)x(s)}} \n \\
        &= \int \cor{D}x(s) \exp{-\frac{1}{2} \int \dd{s_1}\dd{s_2} [x(s_1)-m(s_1)]C^{-1}(s_1,s_2)[x(s_2)-m(s_2)]+ \int \dd{s}j(s)x(s) } \n \\
        &= \exp[\frac{1}{2} \int \dd{s_1}\dd{s_2} j(s_1)C(s_1,s_2)j(s_2) + \int \dd{s}j(s)m(s) ] \; ,
    \label{eq:genfunctional}
\end{align}
where $C(s_1,s_2)$ is given in Eq.~\eqref{eq:2point} and $m(t)$ in Eq.~\eqref{eq:m(t)}. We normalized the integration measure $\cor{D}x(s)$ on the second line of Eq.~\eqref{eq:genfunctional} so that $\cor{Z}[j=0]=1$. We can now use this results to calculate $Q_q(s_1,s_2)$ defined in Eq.~\eqref{eq:Qq_def}. Notice first that, due to the statistical independence of the process along the various spatial coordinates,
\begin{equation}
    Q_q(s_1,s_2) = \prod_{n=1}^d \expval*{ e^{i q_n [X_n^{(0)}(s_2)- X_n^{(0)}(s_1)]} } \; .
\end{equation}
Each of these factors can be simply obtained from $\cor{Z}[j]$ in Eq.~\eqref{eq:genfunctional} by setting $j=j^*(s)\equiv iq_n \left[ \delta(s-s_2)-\delta(s-s_1)\right]$, which yields
\begin{equation}
    Q_q(s_1,s_2) =e^{i \vb{q} \cdot [\vb{m}(s_2)- \vb{m}(s_1)]} e^{-\frac{q^2}{2}\left[ C(s_1,s_1)+C(s_2,s_2)-2C(s_1,s_2) \right] } \; .
    \label{eq:Qq}
\end{equation}
In order to specialize this formula to our problem, let again the particle leave the initial position $\vb{X}_0$ at time $t=t_0$; the effect of having $\vb{X}_0\neq 0$ enters solely in the expression of $\vb{m}(t)$ given in Eq.~\eqref{eq:m(t)}. We may write explicitly, in terms of the two-time function $C(s_1,s_2)$ defined in Eq.~\eqref{eq:2point},
\begin{equation}
e^{-\frac{q^2}{2}\left[ C(s_1,s_1)+C(s_2,s_2)-2C(s_1,s_2) \right] } 
     =
     \begin{cases}
         \exp{-\frac{Tq^2}{k} \left[ 1- e^{-\gamma|s_2-s_1|} - \frac{1}{2}\left( e^{-\gamma s_1} - e^{-\gamma s_2} \right)^2 \right]} & \T{for}\;\; t_0= 0 \; , \\[2mm]
         \exp[-\frac{Tq^2}{k} \left( 1- e^{-\gamma|s_2-s_1|} \right)] & \T{for}\;\; t_0\rightarrow -\infty \; .
     \end{cases}
\label{eq:Qq2}
\end{equation}
In particular, for $t_0=0$,
\begin{equation}
    \vb{m}(s_2)-\vb{m}(s_1) = \vb{X}_0 \left( e^{-\gamma s_2} - e^{-\gamma s_1} \right) \; ,
\end{equation}
while $\vb{m}(t)$ vanishes for $t_0 \rightarrow -\infty$.

In the perturbative calculation of the auto-correlation function discussed further below in Appendix \ref{par:autocorrelation} (see, c.f., Eq.~\eqref{eq:trap_C2}), we also need to derive the expressions for the averages 
\begin{align}
    \expval*{e^{i \vb{q}\cdot \vb{X}\z (t) }} &= \prod_{n=1}^{d} \expval*{e^{i q_n X_n\z (t) }} \; , \\
     \expval*{X_j\z (s_2)e^{i \vb{q}\cdot \vb{X}\z (s_1) }} 
     &= \expval*{X_j\z (s_2) e^{i q_j X_j\z (s_1) }}  \prod_{n \neq j}^{d} \expval*{e^{i q_n X_n\z (s_2) }} \; , \\
    \expval{X_j^{(0)}(s_3) e^{i \vb{q} \cdot [\vb{X}^{(0)}(s_2)- \vb{X}^{(0)}(s_1)]} }
        &= \expval{X_j^{(0)}(s_3)e^{i q_j [X_j^{(0)}(s_2)- X_j^{(0)}(s_1)]}}\prod_{n\neq j}^d \expval{ e^{i q_n [X_n^{(0)}(s_2)- X_n^{(0)}(s_1)]} } \; .
\end{align}
These quantities can be similarly calculated by using the generating functional in Eq.~\eqref{eq:genfunctional}: in fact,
\begin{align}
        \expval*{e^{i q_j X_j\z (t)}} &= \cor{Z} \left[ j(s) = iq_j \delta(s-t)   \right] 
        = e^{-\frac{1}{2} q_j^2 C(t,t)} e^{i q_j m_j(t)} \; , \\
        \expval*{X_j\z (s_2) e^{i q_j X_j\z (s_1) }} &= \eval{\fdv{j(s_2)} \cor{Z}[j]}_{j(s)=iq_j \delta(s-s_1)} 
        = \left[ m_j(s_2) + i q_j C(s_1,s_2) \right] \expval*{e^{i q_j X_j\z (s_1)}} \; , \\
    \expval{X_j^{(0)}(s_3)e^{i q_j [X_j^{(0)}(s_2)- X_j^{(0)}(s_1)]}} &= \eval{\fdv{j(s_3)} \cor{Z}[j]}_{j(s)=iq_j \left[ \delta(s-s_2)-\delta(s-s_1)\right]} \; ,
\end{align}
and we get
\begin{align}
    \expval*{e^{i \vb{q}\cdot \vb{X}\z (t) }} &= e^{-\frac{1}{2} q^2 C(t,t)} e^{i \vb{q} \cdot \vb{m}(t)} \; , \\
    \expval*{\vb{X}\z (s_2)e^{i \vb{q}\cdot \vb{X}\z (s_1) }} &=  \left[ \vb{m}(s_2) + i \vb{q} C(s_1,s_2) \right] \expval*{e^{i \vb{q}\cdot \vb{X}\z (s_1) }} \; , \\
    \expval{\vb{X}^{(0)}(s_3) e^{i \vb{q} \cdot [\vb{X}^{(0)}(s_2)- \vb{X}^{(0)}(s_1)]} } &= \Big\lbrace  \vb{m}(s_3) +i\vb{q} \left[ C(s_2,s_3)-C(s_1,s_3) \right] \Big\rbrace \, Q_q(s_1,s_2)  \; .
    \label{eq:perautocorr}
\end{align}

\section{Equilibrium distribution of the particle}
\label{par:eqcolloid}
The equilibrium distribution of the system composed by the colloidal particle in interaction with the field, the field itself and the thermal bath which provides the thermal noise is given by the Boltzmann distribution
\begin{align}
    P_\T{eq}[\phi,\vb{X}] \propto \exp(-\beta \cor{H}[\phi,\vb{X}]) \; ,
    \label{eq:boltzmann}
\end{align}
where $\beta$ is the inverse temperature of the bath and $\cor{H}$ is the Hamiltonian. Assume that the latter has the generic form
\begin{equation}
    \cor{H}[\phi,\vb{X}] = \cor{H}_\phi[\phi] + \cor{U}(\vb{X}) - \lambda \cor{H}_\T{int}[\phi,\vb{X}] \; ,
    \label{eq:hamiltonian_generic}
\end{equation}
where $\cor{H}_\phi[\phi]$ describes the field in the bulk and is not necessarily Gaussian, while $\cor{U}(\vb{X})$ is a confining potential for the particle, \eg, $\cor{U}(\vb{X}) = (k/2)\vb{X}^2$ in the case considered here. Finally, $\cor{H}_\T{int}$ describes the interaction between the particle and the field via a possibly nonlinear coupling
\begin{equation}
    \cor{H}_\T{int}[\phi,\vb{X}] = \int \dd[d]{x} F[\phi(\vb{x})]V(\vb{x}-\vb{X}) \; ,
    \label{eq:generic_interaction}
\end{equation}
where $F[\phi(\vb{x})]$ is a quasi-local functional of $\phi$; we only require this coupling to be translationally invariant. Our Hamiltonian in Eq.~\eqref{eq:hamiltonian} has indeed the form required in Eq.~\eqref{eq:hamiltonian_generic}. The equilibrium distribution of the colloid follows as
\begin{align}
    P_\T{eq}(\vb{X}) &\propto \int\cor{D}\phi \, e^{-\beta \cor{H}[\phi,\vb{X}]}
    = e^{-\beta \cor{U}(\vb{X})} \int\cor{D}\phi \, e^{-\beta \lgraf \cor{H}_\phi[\phi] - \lambda \cor{H}_\T{int}[\phi,\vb{X}] \rgraf} \; ,
    \label{eq:marginal}
\end{align}
and our aim is to show that the functional integral on the right-hand-side does not actually depend on $\vb{X}$, \ie, that the interaction with the field does not affect the equilibrium distribution $P_\T{eq}(\vb{X}) \propto \exp[ -\beta \cor{U}(\vb{X})]$ of the colloid.
The argument goes as follows: introduce $\vb{z}= \vb{x}-\vb{X}$ and define a new shifted field $\varphi(\vb{z})\equiv \phi( \vb{z}+\vb{X})$. Since the field is in the bulk, then $\cor{H}_\phi[\phi] = \cor{H}_\phi[\varphi]$, while $\cor{H}_\T{int}$ in Eq.~\eqref{eq:generic_interaction} becomes
\begin{equation}
    \cor{H}_\T{int}[\phi,\vb{X}] \rightarrow \int \dd[d]{z} F[\varphi(\vb{z})]V(\vb{z}) \; .
\end{equation}
The proof is concluded by noticing that the integration measure $\cor{D}\phi$ in Eq.~\eqref{eq:marginal} remains the same under a translation by $\vb{X}$ in space.

We emphasize that this argument fails if the system is not translationally invariant, as it happens, for instance, in the presence of boundaries or confinement \cite{gross}. Moreover, it does not imply the factorization of $P_\T{eq}[\phi,\vb{X}]$ into two independent parts at long times. In fact, the marginal equilibrium distribution of the field $\phi$, which may be obtained by integrating out $\vb{X}$ in Eq.~\eqref{eq:boltzmann}, is actually modified by the presence of the colloid. For a linear field-particle coupling such as that of Eq.~\eqref{eq:hamiltonian}, for instance, we physically expect at equilibrium the field to be enhanced around the colloid, \ie, around the minima of its confining potential $\cor{U}(\vb{x})$.

\section{Long-time behavior of the average position (model A and B)}
\label{par:appaverageposition}
In this section we derive the asymptotic behaviour of the second-order correction to the average position in Eq.~\eqref{eq:X2_first} at long times by considering separately the cases of model A and model B field dynamics. By rotational symmetry, we can choose the initial position to have a single non-vanishing coordinate, \ie, $\vb{X}_0(t) = X_0\vb{\hat{j}}$, where $\vb{\hat{j}}$ is the unit vector of the $j$-th Cartesian axis. The resulting average position $\expval*{\vb{X}(t)}$ will then vanish at all times for all but the $j$-th component. Using Eqs.\@~(\ref{eq:Cqfield}), (\ref{eq:field-susc}), (\ref{eq:Qq}), and (\ref{eq:Qq2}), the latter can be written as 
\begin{align}
    \expval*{X_j\t(t)}
    = \frac{\nu}{T} 
        \int & \frac{\dd[d]{q}}{(2\pi)^d} q_j \abs{V_q}^2 \int_0^t\dd{s_2} e^{-\gamma (t-s_2)}
        \int_0^{s_2}\dd{s_1} 
        \left[\alpha_q+\nu T q^2 e^{-\gamma(s_2-s_1)}\right]\times \n \\
        &\times C_q(s_2-s_1)
        \sin\left(q_jX_{0j} (e^{-\gamma s_1}-e^{-\gamma s_2})\right)
        e^{-R(s_2,s_1)q^2} \; ,
        \label{eq:X2-expanded}
\end{align}
where we introduced, for brevity,
\begin{equation}
    R(s_2,s_1) \equiv \frac{T}{k} \left[ 1-e^{-\gamma\abs{s_2-s_1}}-\frac12\left(e^{-\gamma s_2}-e^{-\gamma s_1}\right)^2 \right] \; .
    \label{eq:R(s1,s2)}
\end{equation}
The more general case in which the particle is linearly coupled to the $n$-th even derivative of the field, as it does in Eq.~\eqref{eq:derivativeinteraction}, can be simply accounted for as follows. Note first that Eq.~\eqref{eq:derivativeinteraction} can be rewritten in Fourier space for even $n$ as
\begin{equation}
    \cor{H}_\T{int} = -\lambda \int \dslash{q} \phi_{-q}(t) (iq)^{n} V_{q} e^{-i\vb{q}\cdot \vb{X}(t)} \; .
\end{equation}
It is then enough to replace $V_q$ in Eq.~\eqref{eq:X2-expanded} with $\tilde{V}_q\equiv (iq)^n V_q$.
Since $V(\vb{x})$ is normalized, we can expand the Fourier transform of the rotationally-invariant potential $V_q$ as $|V_q|^2=1 + c_2q^2 + \dots$, whence $|\tilde{V}_q|^2=q^{2n} + c_2q^{2(n+1)} + \dots$. Without loss of generality, $\expval*{X\2_j(t)}$ can then be expressed as a sum of expressions identical to Eq.\@ (\ref{eq:X2-expanded}), but with $q^{2n}$ in place of $|V_q|^2$. Accordingly, in the following we consider the specific case of a potential with $|V_q|^2=q^{2n}$ and we will show that each term in this sum becomes increasingly irrelevant at long times upon increasing $n$.

\subsection{Field with model A dynamics}
\label{par:modelAasymptotics}
In the case of model A dynamics, we start by rescaling $s_1'=s_1/t$, $s_2'=s_2/t$ and $q\to t^{1/2}q$ so as to write Eq.~\eqref{eq:X2-expanded} into the equivalent form 
\begin{align}
    \expval*{X_j\t(t)}
    = t^{-(d/2+n-3/2)}
        \frac{\nu}{T} 
        \int & \frac{\dd[d]{q}}{(2\pi)^d} q_j q^{2n}
        \int_0^1\dd{s_2'} e^{-\gamma t(1-s_2')} \int_0^{s_2'} \dd{s_1'} 
        \left[\alpha_{t^{-1/2}q}+\nu T t^{-1}q^2 e^{-\gamma t(s_2'-s_1')}\right] \times \n\\
        &\times C_{t^{-1/2}q}(t(s_2'-s_1'))
        \sin\left(t^{-1/2}q_jX_{0,j} (e^{-\gamma ts_1'}-e^{-\gamma ts_2'})\right)
        e^{-R(ts_2',ts_1')t^{-1}q^2} \; .
    \label{eq:longtimemodelA}
\end{align}
In this way we removed the time dependence from the integration limits and left it in the integrand only; this is more convenient for considering the limit $t\to\infty$. 
To this end, let us briefly discuss the asymptotic behaviour for $t\to\infty$ of each term in the integrand. The first term in parenthesis tends to
\begin{equation}
    \alpha_{t^{-1/2}q}+\nu t^{-1}T q^2 e^{-\gamma t(s_2'-s_1')} =
    \begin{cases}
        Dr &  \for r>0 \\
        D q^2 t^{-1} & \for r=0
    \end{cases}
    \quad + \quad \ho,
\label{eq:alpha_longtimes}
\end{equation}
where we noticed that in both cases the second addendum is subleading with respect to the first for large $t$. Here and in what follows $h.o.$ denotes additional terms which are subleading in the limit $t\to\infty$. The field correlator tends to
\begin{equation}
    C_{t^{-1/2}q}(t(s_2'-s_1')) =
    \begin{cases}
        Tr^{-1} e^{-D(rt+q^2)(s_2'-s_1')} & \for r>0 \\
        Ttq^{-2}e^{-Dq^2(s_2'-s_1')}      & \for r=0
    \end{cases}
    \quad + \quad \ho \; .
\label{eq:G_longtimes}
\end{equation}
The argument of the sine tends to zero at long times, so that we can expand $\sin x\simeq x$ to leading order. Finally, $R(ts_2',ts_1')t^{-1}q^2$ tends to zero at long times.

Let us now focus on the case $r>0$. The correction to the average position of the particle is then asymptotic to
\begin{align}
    \expval*{X_j\t(t)} &=
    \nu D X_{0,j}C_d
    t^{-(d/2+n-1)} e^{-\gamma t}
    \int_0^\infty\dd{q}
    \int_0^1\dd{s_2'}\int_0^{s_2'}\dd{s_1'} 
    q^{d+2n+1} e^{-D(rt+q^2)(s_2'-s_1')}
    \left[e^{\gamma t(s_2'-s_1')}-1\right]
    \quad+\quad\ho,
\end{align}
where we performed the integration over the angular $q-$variables. The constant $C_d=c_d/d$ comes from the integration of the solid angle in $d$ dimensions, being
\begin{equation}
    c_d \equiv \int \frac{\dd{\Omega_d}}{(2\pi)^d} = \frac{2^{1-d}}{\pi^{d/2} \Gamma(d/2)} \; ,
    \label{eq:solidangle}
\end{equation}
and where we noted that we can replace $q_j^2 \; \rightarrow \; q^2/d$ in the integral.
At this point the integration over $s_2'$ and $s_1'$ can be performed explicitly and we immediately obtain
\begin{align}
    \expval*{X\2_j(t \to \infty)} \propto 
    \begin{cases}
        e^{-\nu kt} & \for \nu k < Dr \; , \\
        e^{-Drt} & \for \nu k > Dr \; .
    \end{cases}
\end{align}
For $r=0$ the asymptotics in Eqs.~\eqref{eq:alpha_longtimes}) and \eqref{eq:G_longtimes} are different from the non-critical case $r>0$, and this affects the asymptotics of Eq.~\eqref{eq:longtimemodelA} which reads
\begin{align}
    \expval*{X_j\t(t)} =
    \nu D^2 X_{0,j}C_d
    t^{-(d/2+n-1)} e^{-\gamma t}
    \int_0^\infty\dd{q}\int_0^1\dd{s_2'}\int_0^{s_2'}\dd{s_1'} 
    q^{d+2n+1} e^{-Dq^2(s_2'-s_1')} 
    \left[e^{\gamma t(s_2'-s_1')}-1\right]
    \quad+\quad\ho \; .
\end{align}
As before, by performing the integration over $s_1'$ and $s_2'$ one gets
\begin{align}
    \expval*{X_j\t(t)} \propto t^{-(d/2+n+1)} \; .
\end{align}
We conclude that, in model A dynamics, an algebraic behavior of the tracer particle is observed at long times only in the critical case $r=0$. These results are summarized in Eq.~\eqref{eq:relax_d_modelA} of the main text.

\subsection{Field with model B dynamics}
The long-time asymptotic behaviour of the average position of the particle in the case of a field with model B dynamics can be found in a similar manner as done above for model A. For $r>0$, we rescale $s_2$, $s_1$ and $q$ as we did in Section \ref{par:modelAasymptotics} in order to obtain Eq.~\eqref{eq:longtimemodelA}. In this case, however, the asymptotic behaviour of the first two terms are
\begin{equation}
    \alpha_{t^{-1/2}q}+\nu T t^{-1}q^2 e^{-\gamma t(s_2'-s_1')}
    = \left[Dr+\nu Te^{-\gamma t(s_2'-s_1')}\right]
    t^{-1}q^2
    \quad+\quad\ho,
\end{equation}
and
\begin{equation}
    C_{t^{-1/2}q}(t(s_2'-s_1')) = Tr^{-1} e^{-Drq^2(s_2'-s_1')}
    \quad+\quad\ho,
\end{equation}
thus leading to
\begin{multline}
    \expval*{X_j\t(t)}
    = \nu X_{0,j} C_d
        t^{-(d/2+n)} e^{-\gamma t}
        \int_0^\infty\dd{q}
        \int_0^1\dd{s_2'} \int_0^{s_2'} \dd{s_1'} \\
        q^{d+2n+3} e^{-Drq^2(s_2'-s_1')}
        \left[e^{\gamma t(s_2'-s_1')}-1\right]
        \left[D+r^{-1}\nu Te^{-\gamma t(s_2'-s_1')}\right]
    \quad+\quad\ho
\end{multline}
As before, the integration over $s_2'$ and $s_1'$ becomes trivial and we obtain
\begin{align}
    \expval*{X_j\t(t)} \propto t^{-(d/2+n+2)} \; .
\end{align}

The case $r=0$ requires, in contrast with the previous ones, that momenta are rescaled in Eq.~\eqref{eq:X2-expanded} as $q\to t^{1/4}q$. In this way we get the equivalent expression
\begin{align}
    \expval*{X_j\t(t)}
    =& t^{-(d/4+n/2-7/4)} \frac{\nu}{T}
        \int\frac{\dd[d]{q}}{(2\pi)^d} q_j q^{2n}
        \int_0^1\dd{s_2'} e^{-\gamma t(1-s_2')} 
        \int_0^{s_2'}\dd{s_1'} 
        \left[
            \alpha_{t^{-1/4}q}+
            \nu T t^{-1/2}q^2 e^{-\gamma t(s_2'-s_1')}
        \right] \times \n\\
        &\times C_{t^{-1/4}q}(t(s_2'-s_1'))
        \sin\left(
            t^{-1/4}q_jX_{0,j} 
            (e^{-\gamma ts_2'}-e^{-\gamma ts_1'})
        \right)
        e^{-R(ts_2',ts_1')t^{-1/2}q^2} \; .
\end{align}
Since for $r=0$
\begin{align}
    \alpha_{t^{-1/4}q}+\nu T t^{-1/2}q^2 e^{-\gamma t(s_2'-s_1')}
        = t^{-1/2}
        \left[
            Dt^{-1/2}q^4 + \nu T e^{-\nu k t(s_2'-s_1')}
        \right]
    \quad+\quad\ho
\end{align}
and
\begin{align}
    C_{t^{-1/4}q}(t(s_2'-s_1'))
        = Tt^{1/2} q^{-2} e^{-Dq^4(s_2'-s_1')}
        \quad+\quad\ho
\end{align}
one has
\begin{align}
    \expval*{X_j\t(t)}
    =& \nu X_{0,j} C_d t^{-(d/4+n/2-3/2)} e^{-\gamma t}
        \int_0^\infty\dd{q} \int_0^1\dd{s_2'} \int_0^{s_2'}\dd{s_1'} 
        q^{d+2n-1} e^{-Dq^4(s_2'-s_1')}
        \left[e^{\gamma t(s_2'-s_1')}-1\right] \times \n\\
       &\times \left[Dt^{-1/2}q^4 + \nu T e^{-\gamma t(s_2'-s_1')}\right]
    \quad+\quad\ho
\end{align}
At this point the integration over $s_1'$ and $s_2'$ is again straightforward and we get
\begin{align}
\expval*{X\2_j(t)} \propto t^{-(d/4+n/2+1)} \; .
\end{align}
We thus conclude that, in model B dynamics, an algebraic behavior of the tracer particle is observed at long times both in the critical case ($r=0$) and off-criticality ($r>0$). These results are summarized in Eq.~\eqref{eq:relax_d} in the main text.

\section{Connection between the long-time behavior of the particle and the critical properties of the field}
\label{par:generalpicture}
We are now in the position to relate the decay exponents of the average particle coordinate which we obtained at criticality to the dynamical critical exponent $z$ of the underlying free-field theory. The key is to introduce the general scaling form of the dynamical susceptibility and two-time function \cite{Tauber}
\begin{align}
    \chi_\phi(q,t) &= \abs{q}^{-2+\eta+z} \chi_\pm \left(q \xi, \frac{\cor{D} a_0^z t}{\xi^z}\right) \; , \label{eq:scaling} \\
    C_\phi(q,t) &= \abs{q}^{-2+\eta} C_\pm\left(q \xi, \frac{\cor{D}a_0^z t}{\xi^z}\right) \; , \label{eq:scaling_corr}
\end{align}
where $\xi$ is the correlation length of the field (which diverges at criticality), $z$ is its dynamical critical exponent and $\eta$ its anomalous dimension; finally, $\cor{D}^{-1}$ and $a_0$ represent some microscopic time and length scales, respectively. The scaling functions $\chi_\pm$ and $C_\pm$ are well-behaved at the critical point, where they take a constant value (depending in general on whether the critical point is approached from above or from below). It is also well-known that $z=2+\alpha$ and $\eta=0$ in the case of model A ($\alpha=0$) and model B ($\alpha=2$) dynamics within the Gaussian approximation \cite{Tauber}.

In order to address the long-time behavior of the average position, we start again from Eq.~\eqref{eq:X2_first} where we identify
\begin{equation}
    \chi_x(t) \equiv \nu \theta(t) e^{-\gamma t}
\end{equation}
as the susceptibility of the particle. We then rescale time as $s=s't$ and momenta as $p=qt^{1/z}$, as suggested by the scaling forms in Eqs.~\eqref{eq:scaling} and \eqref{eq:scaling_corr}. This gives
\begin{align}
    \expval*{X_j\t (t)}  = t^{2-(d+1)/z} &\int \frac{\dd[d]{p}}{(2\pi)^d} i p_j  |V_{pt^{-1/z}}|^2 \int_{0}^1 \dd{s_2'} \int_{0}^{s_2'} \dd{s_1'} \chi_x\left(t(1-s_2')\right) Q_{pt^{-1/z}}\left(ts_1',ts_2'\right) \times \nonumber \\ 
    &\times \left[ \chi_\phi\left(pt^{-1/z},t(s_2'-s_1')\right) +(pt^{-1/z})^2 \chi_x \left(t(s_2'-s_1')\right) C_\phi\left(pt^{-1/z},t(s_2'-s_1')\right) \right] \; ,
    \label{eq:x2_general}
\end{align}
where
\begin{equation}
    Q_q(s_1,s_2) = e^{i \vb{q} \cdot \vb{X_0} \left( e^{-\gamma s_2} - e^{-\gamma s_1} \right)-q^2 R(s_1,s_2)} \; ,
\end{equation}
and $R(s_1,s_2)$ was defined in Eq.~\eqref{eq:R(s1,s2)}. It is easy to check that $p^2 t^{-2/z}R(ts_1',ts_2') \xrightarrow[\gamma t \gg 1]{} 0$, thus
\begin{align}
    &Q_{pt^{-1/z}}(ts_1',ts_2') \simeq 1+i t^{-1/z} \vb{p} \cdot \vb{X}_0 \left( e^{-\gamma ts_2'} - e^{-\gamma ts_1'} \right) \; , \label{eq:expansion_exp} \\
    & i \chi_x\left(t(1-s_2')\right) Q_{pt^{-1/z}}\left(ts_1',ts_2'\right) \simeq \nu t^{-1/z} \vb{p} \cdot \vb{X}_0  \left( e^{-\gamma t(s_2'-s_1')} -1 \right) \; ,
\end{align}
where we omitted an imaginary term from the right hand side of the last equation because it would vanish by symmetry when we integrate over $\vb{p}$ in Eq.~\eqref{eq:x2_general}. The integrand in Eq.~\eqref{eq:x2_general} now only depends on $u\equiv s_2'-s_1'$, so that $v\equiv s_2'+s_1'$ can be integrated out yielding
\begin{align}
    \expval*{X_j\t (t)}  \simeq \nu t^{2-(d+2)/z} e^{-\gamma t} \int \frac{\dd[d]{p}}{(2\pi)^d} p_j  |V_{pt^{-1/z}}|^2 \vb{p} \cdot \vb{X}_0 \int_{0}^1 \dd{u}f(u) \lgraf \chi_\phi\left(pt^{-1/z},tu\right) +(pt^{-1/z})^2 \chi_x \left(tu\right) C_\phi\left(pt^{-1/z},tu\right) \rgraf \; ,
    \label{eq:x2_temp}
\end{align}
where we defined the function
\begin{equation}
    f(u) \equiv (1-u)\left(e^{\gamma t u }-1\right) \; .
\end{equation}
We now look for a saddle-point estimate of the integral over $u$ in Eq.~\eqref{eq:x2_temp}, bearing in mind that we are after terms which can counterbalance the exponential factor $\exp(-\gamma t)$ in front of the integrals, so as to produce an algebraic behavior of $ \expval*{X_j\t (t)}$ for large $t$. We can already drop a subleading term from
\begin{equation}
    f(u) \simeq (1-u)e^{\gamma t u } = \exp{t \left[\gamma u +\frac{1}{t} \ln(1-u)\right]} \equiv e^{tg(u)} \; ,
\end{equation}
where the function $g(u)$ has its maximum in $u^* = 1-(\gamma t)^{-1}$.
The integrands $\chi_\phi$ and $C_\phi$ are both decreasing functions of their second argument and they decay with the relaxation timescale $\tau_\phi$ of the field: we thus expect them not to affect the position of the saddle point whenever $\tau_\phi \gg \tau_X$, \ie, in the presence of slow field modes (with hindsight, we actually know that this argument only fails in model A when we are sufficiently far from criticality so that $Dr>\gamma$). Moreover, due to the additional factor $\chi_x \left(tu\right) \sim e^{-\gamma t u}$ in front, the term containing $C_\phi$ in Eq.~\eqref{eq:x2_temp} is \textit{a priori} subleading for large $t$. We thus obtain
\begin{align}
     &\int_{0}^1 \dd{u}f(u) \left[ \chi_\phi\left( pt^{-1/z},tu\right) +(pt^{-1/z})^2 \chi_x \left(tu\right) C_\phi\left(pt^{-1/z},tu\right) \right] \n \\
     &\simeq  \int_{0}^1 \dd{u}  e^{tg(u)} \chi_\phi\left( pt^{-1/z},tu\right) \simeq \frac{e^{\gamma t-1}}{\gamma t} \chi_\phi\left( pt^{-1/z},t u^*\right) \int_\mathbb{R} \dd{u} \exp[ -\frac{\gamma^2 t^2}{2} (u-u^*)^2  ] \n\\
     &= \frac{\sqrt{2\pi} }{(\gamma t)^2}e^{\gamma t-1} \chi_\phi\left( pt^{-1/z},t-\frac{1}{\gamma}\right) \; ,
\end{align}
from which one reads the general asymptotic result
\begin{align}
    \expval*{X_j\t (t)}  \simeq \frac{ \sqrt{2\pi}\nu}{e \gamma^2}  t^{-(d+2)/z} \int \frac{\dd[d]{p}}{(2\pi)^d} p_j  |V_{pt^{-1/z}}|^2 \left(\vb{p} \cdot \vb{X}_0\right) \chi_\phi\left(pt^{-1/z},t-\frac{1}{\gamma}\right) \; .
    \label{eq:asymptotic_general}
\end{align}
This expression depends on the specific form of the field susceptibility $\chi_\phi(q,t)$ and it is thus in general model-dependent. Close to criticality, however, we can plug in the scaling form Eq.~\eqref{eq:scaling} to get
\begin{align}
    \expval*{X_j\t (t)}  \simeq  t^{-1-(d+2n)/z} \int \frac{\dd[d]{p}}{(2\pi)^d} p_j  p^{2n-2+z} \left(\vb{p} \cdot \vb{X}_0\right)  \chi_\pm\left(pt^{-1/z}\xi,\frac{Da_0^z}{\xi^z} (t-\frac{1}{\gamma})\right) \; ,
    \label{eq:derivedscaling}
\end{align}
where we set $\eta=0$ since we are dealing with a free theory, and we generically assumed $V_q \sim q^n$ for small $q$; this also accounts for the case of a linear coupling to the $n$-th even derivative of the field as in Eq.~\eqref{eq:derivativeinteraction} (see discussion in Appendix \ref{par:appaverageposition}). When $\xi \rightarrow \infty$, the $t$ dependence drops out of the $\dd[d]{p}$ integral and we recover the universal long-time scaling at criticality, see Eq.~\eqref{eq:general_critical}.

Using the free-field susceptibility given in Eq.~\eqref{eq:field-susc} allows to write explicitly, for model A and B at criticality ($r=0$) and within the Gaussian approximation, the asymptotic estimate
\begin{align}
    \expval*{X_j (t)}  \simeq \frac{\sqrt{2 \pi} c_d}{z\, e\, d} \Gamma \left(1+ \frac{d+2n}{z}  \right)\frac{\lambda^2 X_0}{k} \left(\gamma t\right)^{-1} \left(D t\right)^{-(d+2n)/z} \; ,
    \label{eq:asymptotic_critical}
\end{align}
where the constant $c_d$ was defined in Eq.~\eqref{eq:solidangle}, and $\Gamma(x)$ is the Euler Gamma function. This determines the asymptotic amplitude of the average particle position. A similar calculation gives, for the off-critical model B,
\begin{align}
    \expval*{X_j (t)}  \simeq \frac{\sqrt{\pi/2}\, c_d}{e\, d} \Gamma \left(1+ \frac{d+2n}{z}  \right)\frac{\lambda^2 X_0 D}{k\gamma} \left( D r \right)^{-(2+n+d/2)}  t^{-2-(d+n)/2}  \; ,
    \label{eq:asymptotic_massive}
\end{align}
where we noted that the free-field susceptibility simplifies because $(p^2 t^{-2/z}+r)\simeq r$ at long times, and we changed the integration variable to $y=Drp^2 \sqrt{t}$ in Eq.~\eqref{eq:asymptotic_general}.

We emphasize that, in order to derive our expression for the average position of the colloid in Eq.~\eqref{eq:x2_general}, we used explicitly the fact that the Hamiltonian of the field is Gaussian: this makes the equation of motion for $\phi_q(t)$ linear and thus exactly solvable via its linear response propagator. This prevents a direct application of our final scaling result, Eq.~\eqref{eq:derivedscaling}, to an interacting field theory. We postpone the investigation of a possible extension in this direction to future works.

\section{Nonlinear transient behavior for large initial displacements}
\label{app:nonlinear}
In this Appendix we investigate the transient behavior displayed by the perturbative solution in Eq.~\eqref{eq:X2_first} when the initial displacement $X_0$ is chosen sufficiently large so as to depart from the linear response regime. Our analysis is based on the phenomenological observation that a transient regime exists in which the average displacement of the colloid decays algebraically, but with a characteristic intermediate exponent different from the one displayed at longer times. One then observes for some time $t_c$ a crossover to the asymptotic behavior predicted by Eqs.~\eqref{eq:relax_d_modelA} and \eqref{eq:relax_d}; the value of $t_c$ grows as we increase the initial displacement $X_0$. Interestingly enough, the amplitude of the average position in the intermediate regime turns out to be independent of the value of $X_0$ itself. This behavior is well-confirmed by numerical simulations of the system (see, \eg, Fig.~\ref{fig:transient}) and it is already visible at zero temperature; we thus focus here, for simplicity, on the noiseless case.

Consider first the critical case $r=0$. Here the field propagator $\chi_q(t)$ defined in Eq.~\eqref{eq:field-susc} can be expressed, both for model A and B dynamics, in the compact form
\begin{equation}
    \chi_q(t) = Dq^{z-2} e^{-Dq^z t} \theta(t) \; ,
    \label{eq:critical_propagator}
\end{equation}
where $z=2+\alpha$ is the dynamical critical exponent. Stepping to dimensionless variables $s_1 \rightarrow s_1/t$, $s_2 \rightarrow s_2/t$ in Eq.~\eqref{eq:X2_first} and rescaling momenta as $p=q X_0$, we can rewrite our perturbative solution for the average tracer position as \cite{barenblatt}
\begin{align}
    \expval*{X_j\t (t)}  &= \frac{X_0^{1-d}}{k} \Phi \left(\gamma t, \frac{t}{t_c}\right) \; , \label{eq:scalingform} \\
    \Phi(\Pi_1,\Pi_2) &= \Pi_1\Pi_2 \int \frac{\dd[d]{p}}{(2\pi)^d} i p_j p^\alpha \abs{V_{p/X_0}}^2 \int_{0}^1 \dd{s_2} \int_{0}^{s_2} \dd{s_1} e^{-\Pi_1(1-s_2)-\Pi_2 p^z (s_2-s_1)} \exp[ip_j \left( e^{-\Pi_1 s_2} - e^{-\Pi_1 s_1} \right)] \; , \label{eq:scaling_def}
\end{align}
where we set $t_0=0$ and we identified the crossover time $t_c\equiv X_0^z/D$. This has to be compared with the asymptotic expression we found in Eq.~\eqref{eq:general_critical} which, upon setting $n=0$, can be expressed in terms of $t_c$ as
\begin{align}
    \expval*{X_j\t (t)}
 \simeq \frac{c_1 X_0^{1-d}}{k}  \left(\gamma t\right)^{-1}  \left(\frac{t}{t_c}\right)^{-d/z} \;  .
\end{align}
The latter is, in fact, linear in $X_0$, so that at long times we can write
\begin{align}
    \expval*{X_j (t)} \simeq c_\infty X_0 \, t^{-\alpha_\infty} \; ,
 \label{eq:app_asymptotic}
\end{align}
where we introduced $\alpha_\infty \equiv 1+d/z$ and $c_\infty\propto \lambda^2 /(\gamma kD^{d/z}) $ up to a numerical constant (see Eq.~\eqref{eq:asymptotic_critical} in Appendix \ref{par:generalpicture}).

We already noted that the correction in Eq.~\eqref{eq:X2_first} to the average position vanishes at time $t=0$ (as well as for $t\rightarrow \infty$), and thus the function $\Phi(\Pi_1,\Pi_2)$ vanishes for $\Pi_1=\Pi_2=0$. However, studying such function analytically is difficult, mostly because $\Pi_1$ and $\Pi_2$ cannot really be treated as independent variables. Notice, moreover, that a residual dependence on $X_0$ is left into the integral over the variable $p$ in Eq.~\eqref{eq:scaling_def} even after introducing dimensionless variables, thus complicating the analysis even further. Some progress can be made by assuming that, when $t_c \gg \tau_X= \gamma^{-1}$ and the leading order $\order{\lambda^0}$ exponential term has become negligible, the average position of the colloid evolves according to a different scaling form, namely Eq.~\eqref{eq:scaling_transient} of the main text. This second ansatz incorporates the phenomenological observation that the amplitude of the average position is independent of $X_0$ within the transient region $t\lesssim t_c$, while the behavior as a function of time $t$ remains algebraic with an exponent $\alpha_0\neq \alpha_\infty$. The underlying physical intuition is the following. At time $t=0$ the colloid is put in contact with the field at position $X(0)=X_0$, and at short times it is dragged primarily by the restoring force of the harmonic trap, so that $\dot{X}\simeq -\gamma X_0$. On a timescale given by $\tau_X=\gamma^{-1}$ the colloid covers a distance $\Delta X$ of the order of $\Delta X \sim X_0$, so that it becomes relevant to take into account the time $t_c(X_0)$ taken by the field to rearrange over such a distance. This allows us to identify $t_c(X_0) \equiv \tau_\phi(q\sim1/X_0)$ in the language of Eq.~\eqref{eq:tau_phi}, which tells us in particular that when $r=0$ this timescale is given by $t_c= X_0^z/D$. When $X_0$ is small, on the other hand, we enter the regime in which $t_c \ll \tau_X$ and no crossover is observed within the asymptotic region $t\gg\tau_X$. 

The intermediate algebraic decay exponent $\alpha_0$ can be determined by comparing the asymptotic form of Eq.~\eqref{eq:scaling_transient} with Eq.~\eqref{eq:app_asymptotic} and by matching powers of $X_0$ in the two expressions. This gives $\beta_0=1/z$, and thus
\begin{equation}
    \alpha_0 = \alpha_\infty-\beta_0 = 1+\frac{d-1}{z} \; .
    \label{eq:intermediate_exponent}
\end{equation}
This matching additionally instructs on which parameters control the amplitude of the average particle position within the transient regime, yielding, up to some numerical constant,
\begin{equation}
    c_0 \propto D^{1/z} c_\infty \propto \frac{\lambda^2}{\gamma k}D^{(1-d)/z} \; .
\end{equation}
We verified that Eq.~\eqref{eq:intermediate_exponent} correctly predicts the intermediate exponent in numerical plots of Eq.~\eqref{eq:X2_first} and in numerical simulations of the system in various spatial dimensions $d$. Alternatively, we can rephrase Eq.~\eqref{eq:scaling_transient} as
\begin{equation}
     \expval*{X_j(t)} \simeq c_0 t_c^{\alpha_0} f_2\left( \tau \right) \; ,
\end{equation}
which is a function of the parameter $\tau=t/t_c$ only, having identified $f_2(\tau) \equiv \tau^{-\alpha_0} f(\tau)$. In the inset of Fig.~\ref{fig:transient} we thus plot $ t_c^{-\alpha_0} \expval*{X_j(t)} \propto  f_2\left( \tau \right)$ in order to observe this scaling function from the collapse of numerical curves obtained for different values of the initial displacement $X_0$.

Above we presented the argument for the critical case $r=0$, but it can be easily extended so as to cover the off-critical case in model B, which also displays an eventual algebraic decay. From Eq.~\eqref{eq:tau_phi} we read the crossover time
\begin{equation}
    t_c \sim \frac{X_0^2}{D r} \; ,
\end{equation}
valid for $r^{-1/2}=\xi\ll X_0$. The asymptotic matching of Eq.~\eqref{eq:scaling_transient} with the long-time decay exponents given in Eq.~\eqref{eq:relax_d} for $r>0$, \ie, $\alpha_\infty=2+d/2$, yields the prediction $\alpha_0=2+(d-1)/2$ for the intermediate decay exponent, and
\begin{equation}
    c_0 \propto \sqrt{D r} c_\infty \propto \frac{\lambda^2 D}{\gamma k}\left(D r \right)^{-(d+3)/2} \; .
\end{equation}
This correctly describes the transient behavior observed in numerical simulations for the off-critical case in model B.

\section{Long-time behavior of the auto-correlation function of the particle}
\label{par:autocorrelation}
The auto-correlation function of the probe in the stationary state can be expanded in powers of the coupling constant $\lambda$ as
\begin{align}
    \expval*{\vb{X}(t)\cdot \vb{X}(t')} = C_0(t-t') + \lambda^2 C_2(t-t') + \order{\lambda^4} \; ,
\end{align}
where the second-order correction reads \cite{wellGauss}
\begin{equation}
    \label{eq:trap_C2}
    C_2(t-t') = 
        \expval*{\vb{X}\1(t)\cdot \vb{X}\1(t')} +
        \expval*{\vb{X}\0(t)\cdot \vb{X}\2(t')} +
        \expval*{\vb{X}\0(t')\cdot \vb{X}\2(t)} \; .
\end{equation}
In this equation and in the following, the quantities $\vb{X}^{(n)}(t)$ are computed in the same way as discussed in Sec.~\ref{par:Relaxation}. With some straightforward calculations one obtains
\begin{align}
    \expval*{\vb{X}\1(t)\cdot \vb{X}\1(t')}
        =& \nu^2 c_d\int_{-\infty}^t\dd{s_2}
        e^{-\gamma(t-s_2)}
        \int_{-\infty}^{t'}\dd{s_1}
             e^{-\gamma(t'-s_1)}
        \int_0^\infty\dd{q} q^{d+1} \abs{V_q}^2
            Q_q^\text{eq}(s_2-s_1) C_q(s_2-s_1) \; ,
    \label{eq:autocorr1}\\
    \expval*{\vb{X}\2(t)\cdot \vb{X}\0(t')}
    =& \frac{\nu c_d}{k} 
        \int_{-\infty}^t \dd{s_2} e^{-\gamma(t-s_2)}
        \int_{-\infty}^{s_2} \dd{s_1}
        \big(e^{-\gamma \abs{t'-s_1}} - e^{-\gamma \abs{t'-s_2}}\big) \times \n \\
        &\times \int_0^\infty\dd{q}q^{d+1}
        \abs{V_q}^2 Q_q^\text{eq}(s_2-s_1) C_q(s_2-s_1)
        \left[
            \alpha_q + \nu Tq^2 e^{-\gamma(s_2-s_1)}
        \right] \; .
    \label{eq:autocorr2}
\end{align}
In these expression we introduced
\begin{align}
    Q^\text{eq}_q(t-t') 
    =\exp[-\frac{Tq^2}{k}(1-e^{-\gamma \abs{t-t'}})] \; ,
    \label{eq:Q_eq}
\end{align}
which is the equilibrium expression of $Q_q(t,t')$ in Eq.~\eqref{eq:Qq}, obtained by taking its limit for $t_0\to - \infty$. In order to derive Eqs.~\eqref{eq:autocorr1} and \eqref{eq:autocorr2}, one also needs the stationary value of the average defined in Eq.~\eqref{eq:perautocorr}, \ie,
\begin{align}
    \expval*{\vb{X}\z(s) e^{i\vb{q}\cdot(\vb{X}\z(t)-\vb{X}\z(t'))}}
    \xrightarrow[t_0 \to -\infty]{} \frac{T}{k}\left[e^{-\gamma \abs{s-t}} -e^{-\gamma\abs{s-t'}}\right]
    i\vb{q}Q_q^{\T{eq}}(t-t') \; .
\end{align}
Finally, the constant $c_d$ comes from the integration of the angular variables and it is defined in Eq.~\eqref{eq:solidangle}.

The procedure to obtain the long-time expansion of the autocorrelation function is very similar to the one used for the average position in Appendix \ref{par:appaverageposition} and will be briefly summarized here. For simplicity, let us set $t'=0$. In order to expand $\expval*{\vb{X}\1(t)\cdot \vb{X}\1(0)}$ for large $t$, one starts by rescaling $s_2\to s_2/t$, $s_1\to s_1/t$ and $q\to t^{1/2}q$ for model A and non-critical model B, while $q\to t^{1/4}q$ for critical model B. Then one proceeds by expanding the resulting integrands for long times. Integrating over $s_2$, $s_1$, one finds for $\expval*{\vb{X}\1(t)\cdot \vb{X}\1(0)}$ the asymptotic behavior reported in Eqs.~\eqref{eq:corr_d_modelA} and \eqref{eq:corr_d} of the main text.

The determination of the asymptotic behaviour of the other two terms in Eq.~\eqref{eq:trap_C2} is slightly more involved due to the increased complexity of Eq.~\eqref{eq:autocorr2}. It actually turns out that the second term in Eq.~\eqref{eq:trap_C2} is of the same order as the first term in the case of model A dynamics, while it is subleading for model B; the third term in Eq.~\eqref{eq:trap_C2} is, instead, always subleading with respect to the first. These facts are proven in Appendix A of Ref.~\cite{FerraroThesis}. With these observations, the asymptotic long-time behavior of $\expval*{\vb{X}(t)\cdot \vb{X}(0)}$ is then the same as that of $\expval*{\vb{X}\1(t)\cdot \vb{X}\1(0)}$. These results coincide with those of Ref.~\cite{wellGauss} for the case of model B.

Finally, as a check we show that the correction in Eq.~\eqref{eq:trap_C2} to the variance $\expval*{X^2(t)}$ vanishes at long times $t$: indeed, the equilibrium distribution of the colloid cannot depend on the coupling $\lambda$ (see Appendix \ref{par:eqcolloid}). Setting $t'=t$ in Eq.~\eqref{eq:autocorr1}, calling $u\equiv s_2-s_1$ and integrating over $v\equiv 2t-(s_2+s_1)$ gives for the first contribution
\begin{align}
    \expval*{\vb{X}\1(t)\cdot \vb{X}\1(t)}
        = \frac{\nu c_d}{k} \int_0^\infty \dd{u} e^{-\gamma u} \int_0^\infty \dd{q} q^{d+1} \abs{V_q}^2 Q_q^\text{eq}(u) C_q(u) \; ,
        \label{eq:app_x1x1}
\end{align}
which correctly does not depend on $t$ (the stationary state is time-translational invariant). The second contribution can be worked out by again changing variables as $s_1 \rightarrow u\equiv s_2-s_1$ and by noticing that
\begin{equation}
    \dv{u} \left[ Q_q^\text{eq}(u) C_q(u) \right]
    = - Q_q^\text{eq}(u) C_q(u) \left[
            \alpha_q + \nu Tq^2 e^{-\gamma u}
        \right] \; ,
\end{equation}
which can be proved by direct inspection of Eqs.~\eqref{eq:field_corr} and \eqref{eq:Q_eq}. Integrating in $u$ by parts and evaluating the integral over $s_2$ finally gives
\begin{align}
    \expval*{\vb{X}\2(t)\cdot \vb{X}\0(t)}
        = - \frac{\nu c_d}{2k} \int_0^\infty \dd{u} e^{-\gamma u} \int_0^\infty \dd{q} q^{d+1} \abs{V_q}^2 Q_q^\text{eq}(u) C_q(u) \; .
\end{align}
Taking into account this explicit form together with Eq.~\eqref{eq:app_x1x1}, we deduce that the sum of all the contributions in Eq.~\eqref{eq:trap_C2} vanishes, as it should.

\section{Adiabatic elimination of the field degrees of freedom}
\label{par:adiabatic}
Following Refs.~\cite{risken,kaneko}, we derive an adiabatic approximation of the dynamics of the system which consists in integrating out the field degrees of freedom from the Fokker-Planck equation \eqref{eq:fpadiabatic} under the assumption that they relax much faster than the position $\vb{X}(t)$ of the particle. As we have denoted by $D$ and $\nu$ the mobility of the field and the colloid respectively, which set the timescales for their relaxation, we will use their ratio $\nu/D$ as a small parameter for our expansion. A later comparison with the weak-coupling solution discussed in Section \ref{par:Relaxation} will lead us to conclude that the adiabatic approximation is in fact only reliable for a dissipative field dynamics (model A) and sufficiently far from the critical point so that $Dr\gg \gamma\equiv \nu k$, being $\tau_X = \gamma^{-1}$ the relaxation timescale of the colloid.

To simplify the notation, let us first rewrite the two coupled Langevin equations \eqref{eq:coll_adiabatic} and \eqref{eq:decoupled} as
\begin{align}
          \dot{\vb{X}}&= -\nu k \vb{X} + \sum_{\sigma=R,I}  \int_{\mathbb{R}^d} \dslash{q} \vb{f}_q^\sigma \phi_q^\sigma + \bm{\xi}(t) \; \equiv \;  \vb{F}(\vb{X},\phi;t) + \bm{\xi}(t) \; ,
    \label{eq:X_appendix} \\
        \dot{\phi}_q^\sigma &= -\alpha_q \phi_q^\sigma +D\lambda q^\alpha g_q^\sigma  + \zeta_q^\sigma  \; \equiv \;  -a \phi_q^\sigma - b + \zeta_q^\sigma \; ,
\end{align}
where $\sigma=R,I$ indicate the real or the imaginary part respectively. Here we introduced $g_q(\vb{X}) \equiv V_q \exp(-i\vb{q}\cdot \vb{X})$ as in Section \ref{par:effective_FP}, while
\begin{equation}
    \begin{cases}
    a \equiv \alpha_q = Dq^\alpha(q^2+r) \; , \\
    b \equiv -D \lambda q^\alpha g_q^\sigma \;  ,\\
    c \equiv \Gamma_\phi/2 = DTq^\alpha \; ,
    \end{cases}
    \label{eq:abc}
\end{equation}
and $\vb{f}_q^\sigma \equiv \nu \lambda \vb{q} \mqty(g_q^I \\ -g_q^R)$. The corresponding noise amplitudes can be simply obtained from Eqs.~\eqref{eq:part_noise} and \eqref{eq:field_noise},
\begin{align}
    \expval*{\xi_{i}(t) \xi_{j}(t') } &= 2\nu T \delta_{ij}\delta(t-t') \equiv \Gamma_x  \delta_{ij}\delta(t-t') \; , \\
    \expval*{\zeta_q^\sigma (t)\zeta_{q'}^\sigma(t')}&=  \frac{\Gamma_\phi}{2} \left[\delta^d(q-q') \pm \delta^d(q+q')\right]\delta(t-t') \; , \label{eq:app_noise_corr}
\end{align}
where the plus sign in Eq.~\eqref{eq:app_noise_corr} is taken for the real part $\sigma=R$ and the minus sign for $\sigma=I$. The average values of all the noises involved here vanish, and so do the cross correlations such as $\expval*{\zeta_q^R (t)\zeta_{q'}^I(t')}$ or $\expval*{\zeta_q^\sigma (t) \xi_{j}(t')}$.

We notice that rescaling time in Eq.~\eqref{eq:X_appendix} as $t\rightarrow \nu t$ is tantamount to setting $\nu\equiv 1$ and replacing $D \rightarrow \widetilde{D} = D/\nu$ in all the above relations. We will henceforth use $\widetilde{D}^{-1}$ as an adiabaticity parameter. 

In this notation, the Fokker-Planck equation for the joint probability distribution $\cor{P}\left[\phi,\vb{X},t\right]$ becomes
\begin{equation}
    \partial_t \cor{P} = \left[ \cor{L}_X + \sum_{\sigma=R,I}  \int_{\mathbb{R}^d} \dslash{q} \cor{L}_q^\sigma \right] \cor{P} \; ,
    \label{eq:fp2adiabatic}
\end{equation}
where, recalling the definition of $\vb{F}$ in Eq.~\eqref{eq:X_appendix}, we introduced 
\begin{equation}
    \cor{L}_X = - \div \vb{F}  + \frac{\Gamma_x}{2} \laplacian \; ,
\end{equation}
and, calling $\partial_\phi \equiv\fdv{\phi_q^\sigma}$,
\begin{eqnarray} 
    \cor{L}_q^\sigma = \partial_\phi (a \phi +b) +c \partial_\phi^2 \; .
    \label{eq:Lq}
\end{eqnarray}
Let us also denote, for brevity,
\begin{equation}
     \int^{'} \dd[d]{q} \equiv \sum_{\sigma=R,I}  \int_{\mathbb{R}^d} \dslash{q} \; ,
\end{equation}
and omit the indication of the superscript $\sigma$ from now on: a further dependence on $\sigma$ will be understood whenever a quantity depends on $q$.

The approach described below resembles closely the Born-Oppenheimer approximation for solving the Schrödinger equation for an atom under the assumption that the nucleus dynamics is much slower than that of the surrounding electrons. 
For each fixed $\vb{X}$, we consider the eigenfunctions $\varphi_{n_q}(\phi_q;\vb{X})$ of the operators $\cor{L}_q$ defined in Eq.~\eqref{eq:Lq}, each satisfying an eigenvalue equation
\begin{equation}
    \lambda_{n_q}(\vb{X})\varphi_{n_q}(\phi_q;\vb{X}) = - \cor{L}_q \varphi_{n_q}(\phi_q;\vb{X}) \; .
    \label{eq:eigenvalue}
\end{equation}
We can expand the joint probability density $\cor{P}\left[\phi,\vb{X},t\right]$ as
\begin{equation}
    \cor{P}\left[\phi,\vb{X},t\right] = \sum_{\vb{n}} P_{\vb{n}}(\vb{X},t)\Phi_{\vb{n}}\left[\phi;\vb{X}\right]
    \label{eq:expansion}
\end{equation}
where $\vb{n}=\{ n_q \}$ is the collection of the excitation numbers for each mode, and we introduced
\begin{equation}    
    \Phi_{\vb{n}}\left[\phi;\vb{X}\right] \equiv \prod_{q \in \mathbb{R}^d}{\vphantom{\prod}}' \varphi_{n_q}\left( \phi_q;\vb{X} \right) 
\end{equation}
(the prime sign again indicates a further product over real and imaginary parts). In particular, using the property \cite{risken}
\begin{equation}
    \int \cor{D} \phi \,  \Phi_{\vb{n}}\left[\phi;\vb{X}\right] = \delta_{\vb{n}\vb{0}} \; ,
    \label{eq:normalization}
\end{equation}
one can show that the marginal probability distribution $P_0(\vb{X},t)$ of the position of the particle can be obtained as
\begin{equation}
    P_0(\vb{X},t) = \int \cor{D}\phi \, \cor{P}\left[\phi,\vb{X},t\right] \; .
    \label{eq:marginal2}
\end{equation}
In the following, we will thus derive an effective evolution equation for $P_0(\vb{X},t)$.

\subsection{Transformation to a Schrödinger-type operator}
\label{par:transformation}
It is well-known (see, \eg, Ref.~\cite{risken}) that a Fokker-Planck operator $\cor{L}_\T{FP}$ acting on a probability distribution $P$ can be brought, under suitable conditions, into a self-adjoint form via a similarity transformation
\begin{equation}
    \cor{H}_\T{FP} = e^{\Phi_\T{st}/2}\cor{L}_\T{FP} e^{-\Phi_\T{st}/2} \; .
    \label{eq:similarity}
\end{equation}
In the case of natural boundary conditions \cite{risken}, the function $\Phi_\T{st}$ is simply related to the stationary distribution $P_\T{st} = \cor{N} \exp{-\Phi_\T{st}}$, 
where $\cor{N}$ is a normalization constant. This way the Fokker-Planck equation
\begin{equation}
    \partial_t P = \cor{L}_\T{FP} P
\end{equation}
takes the form of a time-dependent Schrödinger equation in imaginary time for the transformed $\widetilde{P}=e^{\Phi_\T{st}/2} P$,
\begin{equation}
    \partial_t \widetilde{P} = \cor{H}_\T{FP} \widetilde{P} \; .
\end{equation}
One can check that its eigenfunctions $\psi_n$ defined by
\begin{equation}
    \varphi_n = \psi_n \psi_0 \; , \;\;\; \T{with} \;\;\; \psi_0^2 = P_\T{st} \; ,
    \label{eq:new_eigenfunctions}
\end{equation}
have the same eigenvalues as $\cor{L}_\T{FP}$ and form an orthonormal set,
\begin{equation}
    \int \psi_n \psi_m = \delta_{nm} \; .
    \label{eq:orthogonality}
\end{equation}
Now we observe that each of the operators $\cor{L}_q^\sigma$ defined in Eq.~\eqref{eq:Lq} can be mapped onto
\begin{equation}
    \cor{H}_\T{FP} = c\partial_\phi^2 - W(\phi) \equiv -a \hat{a}^\dag \hat{a}
\end{equation}
using the similarity transformation in Eq.~\eqref{eq:similarity}. Here 
\begin{equation}
    W(\phi) = \frac{a^2}{4c}\left(\phi+\frac{b}{a}\right)^2 -\frac{a}{2}
\end{equation}
is a simple harmonic potential, while $\hat{a}^\dag$ and $\hat{a}$ are the raising and lowering bosonic operators defined by
\begin{equation}
    \hat{a} = \sqrt{\frac{a}{4c}}\left( \tilde{\phi} + \frac{2c}{a} \partial_{\tilde{\phi}} \right)\;, \;\;\;\;\;  \hat{a}^\dag = \sqrt{\frac{a}{4c}}\left( \tilde{\phi} - \frac{2c}{a} \partial_{\tilde{\phi}} \right) \; ,
\end{equation}
where we introduced
\begin{equation}
    \tilde{\phi} \equiv \phi + \frac{b}{a} = \sqrt{\frac{c}{a}} \left( \hat{a} +\hat{a}^\dag \right) \; .
    \label{eq:tilde_phi}
\end{equation}
The solution to the eigenvalue problem (analogous to Eq.~\eqref{eq:eigenvalue})
\begin{equation}
    \lambda_n \psi_n = - \cor{H}_\T{FP} \psi_n
\end{equation}
is then simply given by the set of eigenvalues $\lambda_n = a\,n $, and the corresponding eigenfunctions are
\begin{align}
    \psi_n(\tilde{\phi}) &= \frac{1}{\sqrt{2^n n!}}H_n \left( \sqrt{\frac{a}{2c}}\tilde{\phi} \right) \psi_0(\tilde{\phi}) \; , \\
    \psi_0(\tilde{\phi}) &= \left( \frac{a}{2\pi c} \right)^{1/4} e^{-a \tilde{\phi}^2 / (4c)} \; ,
\end{align}
where $H_n(z)$ are Hermite polynomials \cite{arfken}. Note that using $P_\T{st}=\psi_0^2(\tilde{\phi})$, as prescribed by Eq.~\eqref{eq:new_eigenfunctions}, correctly renders the $(q,\sigma)$-dependent part of the stationary distribution of the field at fixed colloid position $\vb{X}$, which is in our case the canonical one. Indeed, rewriting the Hamiltonian in Eq.~\eqref{eq:hamiltonian} in Fourier space we get
\begin{align}
        \cor{P}_\T{st}(\phi;x) &\propto e^{-\beta \cor{H}}
        \propto \exp{-\beta \int \dslash{q}  \left[ \frac{1}{2}(q^2+r) \phi_q\phi_{-q}-\lambda V_q e^{-i\vb{q}\cdot\vb{X}} \phi_q \right] } \n \\
        &= \exp{-\beta \int \dslash{q}  \left[ \frac{1}{2}(q^2+r)\left((\phi_q^R)^2+(\phi_q^I)^2\right)-\lambda \left( \phi_q^R g_q^R +\phi_q^I g_q^I\right) \right] } \; ,
\label{eq:stationary}
\end{align}
which factorizes over the modes and their real and imaginary parts. This coincides with $P_\T{st}=\psi_0^2(\tilde{\phi})$ upon substituting the definition of $\tilde{\phi}$ in Eq.~\eqref{eq:tilde_phi} and of $a$, $b$, $c$ in Eq.~\eqref{eq:abc}.

\subsection{Effective equation}
Following Ref.~\cite{kaneko}, we now generalize that approach from two to an infinite set of coupled Langevin equations. In order to obtain an evolution equation for each of the $P_{\vb{n}}(\vb{X},t)$, we first substitute the expansion of $ \cor{P}\left[\phi,\vb{X},t\right]$ in Eq.~\eqref{eq:expansion} into the Fokker-Planck equation \eqref{eq:fp2adiabatic} and we multiply both sides by $\Psi_{\vb{m}}/\Psi_0$, where 
\begin{equation}    
    \Psi_{\vb{m}}\left[\phi;\vb{X}\right] = \prod_{q \in \mathbb{R}^d}{\vphantom{\sum}}' \psi_{m_q}\left( \phi_q;\vb{X} \right)
\end{equation}
(recall $\Phi_{\vb{m}} = \Psi_{\vb{m}}\Psi_{0}$). Then we take the functional integral over $\cor{D}\phi$ and use the orthogonality relation
\begin{equation}
    \int \cor{D}\phi \Psi_{\vb{m}} \Psi_{\vb{n}} = \delta_{\vb{m}\vb{n}} \; ,
\end{equation}
which follows from Eq.~\eqref{eq:orthogonality}. We now notice that the eigenvalue equation \eqref{eq:eigenvalue} implies
\begin{equation}
       \int^{'} \dd[d]{q} \cor{L}_q  \Phi_{\vb{n}}\left[\phi;\vb{X}\right] = - \int^{'}  \dd[d]{q} \lambda_{n_q}[\vb{X}]\, \Phi_{\vb{n}}\left[\phi;\vb{X}\right] 
       = -\lambda_{\vb{n}} \Phi_{\vb{n}}\left[\phi;\vb{X}\right] \; ,
\end{equation}
where we introduced
\begin{equation}
    \lambda_{\vb{n}} \equiv \int^{'} \dd[d]{q} \lambda_{n_q} = \int^{'} \dd[d]{q} n_q a_q\; .
\end{equation}
Some straightforward algebra \cite{kaneko} then gives (omitting the various functional dependencies from $P_{\vb{n}}=P_{\vb{n}}(\vb{X},t)$)
\begin{equation}
    \partial_t P_{\vb{m}} = \sum_{\vb{n}} \expval*{\frac{\Psi_{\vb{n}}}{\Psi_{\vb{0}}} \cor{L}_X \Phi_{\vb{n}} } P_{\vb{n}} - \lambda_{\vb{m}} P_{\vb{m}} \; ,
    \label{eq:dt_P}
\end{equation}
where by the average symbol we mean
\begin{equation}
    \expval*{\cdots} = \int \cor{D}\phi \, (\cdots) = \int \prod_{q \in \mathbb{R}^d}{\vphantom{\sum}}' \dd{\phi_q} (\cdots) \; .
\end{equation}
In particular, the marginal distribution for the particle position defined in Eq.~\eqref{eq:marginal2} evolves according to
\begin{equation}
    \partial_t P_{\vb{0}} = \expval*{ \cor{L}_X \Phi_{\vb{0}} } P_{\vb{0}} + \sum_{\vb{n} \neq \vb{0}} \expval*{ \cor{L}_X \Phi_{\vb{n}} } P_{\vb{n}} 
    \label{eq:dt_P_0}
\end{equation}
because $\lambda_{\vb{0}}=0$ (we proved this in Sec.~\ref{par:transformation}). By $\vb{n}\neq 0$ in the sum above we mean that the numbers $n_q$ cannot be all zero simultaneously. Since all the other $P_{\vb{m}}$'s decay on timescales $\lambda_{\vb{m}}^{-1} \propto \widetilde{D}^{-1}$, we can solve for $P_{\vb{m}}$ up to $\order{\widetilde{D}^{-1}}$ by setting $\partial_t P_{\vb{m}} = \delta_{\vb{m}\vb{0}}$ in Eq.~\eqref{eq:dt_P} and keeping only the term $\vb{n}=0$ in the sum, whence
\begin{equation}
    P_{\vb{m}} = \frac{1}{\lambda_{\vb{m}}} \expval*{\frac{\Psi_{\vb{m}}}{\Psi_{\vb{0}}} \cor{L}_X \Phi_{\vb{0}} } P_{\vb{0}} + \order{\frac{1}{\widetilde{D}^2}} \; .
\end{equation}
Plugging this result back into Eq.~\eqref{eq:dt_P_0} and using the fact that $\expval*{\laplacian \Phi_{\vb{n}}} = 0 $ because of Eq.~\eqref{eq:normalization}, we get an evolution equation for the reduced probability density 
\begin{equation}
    \partial_t P_0(\vb{X},t) = \cor{L}^{\T{eff}} P_0(\vb{X},t) \; ,
    \label{eq:reduced}
\end{equation}
where
\begin{align}
         &\cor{L}^{\T{eff}} = -\div \expval*{\vb{F} \Psi_0^2} + \frac{\Gamma_x}{2} \laplacian   \label{eq:kaneko} \\
         &+\sum_{{\vb{n}}\neq 0} \div \expval*{\Psi_{\vb{n}} \vb{F} \Psi_0}\frac{1}{\lambda_{\vb{n}}} \left[ \div \expval*{\Psi_{\vb{n}} \vb{F} \Psi_0} - \expval*{\Psi_0^2 \vb{F}\cdot \left( \grad \frac{\Psi_{\vb{n}}}{\Psi_0}\right)} \right. 
         \left. - 2\Gamma_x \expval*{\Psi_{\vb{n}} \grad \Psi_0}\cdot \grad -\frac{\Gamma_x}{2} \expval*{\frac{\Psi_{\vb{n}}}{\Psi_0} \left( \laplacian \Psi_0^2 \right)} \right] +\order{\frac{1}{\widetilde{D}^2}} \; . \n
\end{align}
Recall that in general $\lambda_{\vb{n}}= \lambda_{\vb{n}}[\vb{X}]$ (although not in our specific case), while we did not indicate the dependence on $\vb{X}$ and $\phi$ of $\vb{F}$ and $\Psi_{\vb{n}}$ so as to lighten the notation. The result in Eq.~\eqref{eq:reduced} is analogous to Eq.~(2.13) in Ref.~\cite{kaneko}, which was derived for the simple case of two coupled scalar equations (notice that the analog of the second term in braces in Eq.~\eqref{eq:kaneko} is reported with the wrong sign in Ref.~\cite{kaneko}).

In order to compute the averages which appear in Eq.~\eqref{eq:kaneko}, we can make explicit use of the fact that $\vb{F}$ is linear in each of the $\phi_q$, hence also in the bosonic creation and annihilation operators $\hat{a}_q^\dag$ and $\hat{a}_q$. Let us inspect explicitly one of these terms:
\begin{align}
        \expval*{\Psi_{\vb{n}} \vb{F} \Psi_0} &= -\nu k \vb{X} \expval*{\Psi_{\vb{n}} \Psi_0} +\int \dslash{q} \vb{f}_q \mel{\Psi_{\vb{n}}}{\tilde{\phi}_q -\frac{b_q}{a_q}}{\Psi_0} \n \\
        &= -\nu k \vb{X} \delta_{\vb{n}0} +\int \dslash{q} \vb{f}_q \left[ \sqrt{\frac{c_q}{a_q}} \mel{\Psi_{\vb{n}}}{ \hat{a}_q +\hat{a}_q^\dag }{\Psi_0} -\frac{b_q}{a_q}
        \expval*{\Psi_{\vb{n}} \Psi_0} \right] \n \\
        &= -\nu k \vb{X} \delta_{\vb{n}0} +\int \dslash{q} \vb{f}_q \left[ \sqrt{\frac{c_q}{a_q}} \delta_{\vb{n}q} -\frac{b_q}{a_q} \delta_{\vb{n}0} \right] \; ,
\end{align}
where we used the definition of $\tilde{\phi}_q$ in Eq.~\eqref{eq:tilde_phi}; the $\delta_{\vb{n}q}$ in the last line selects the element $\vb{n}$ with $n_p =\delta_{pq}$, while $\delta_{\vb{n}0}$ selects the "ground state" with $n_p=0$ for every $p$. Next, note that 
\begin{equation}
    \div \expval*{\Psi_{\vb{n}} \vb{F} \Psi_0}\frac{1}{\lambda_{\vb{n}}}  \div \expval*{\Psi_{\vb{n}} \vb{F} \Psi_0} 
    = \sum_{i,j} \partial_i \expval*{\Psi_{\vb{n}} F_i \Psi_0}\frac{1}{\lambda_{\vb{n}}} \partial_j  \expval*{\Psi_{\vb{n}} F_j \Psi_0} \; ,
\end{equation}
and for the sake of simplicity we will assume isotropy of the interaction potential (\ie, $V_q$ depends only on $q=|\vb{q}|$). This implies that no mixed derivative of the form $\partial_i \partial_j$ will survive the $\dd[d]{q}$ integration, so that a $\delta_{ij}$ can be understood in the sum. Similar considerations apply to the other averages in Eq.~\eqref{eq:kaneko}, which can be dealt with using the properties of Hermite polynomials (or equivalently the bosonic algebra); it is crucial at some point to reinstate the dependence on $\sigma$, because many of the contributions cancel out when taking $\sum_\sigma$. A lengthy but simple computation gives, once reinstating the original parameters $\nu$ and $D$,
\begin{equation}
    \cor{L}^{\T{eff}} = \sum_{j=1}^d  \left[ \partial_j \left( \chi_j \nu k X_j \right) +  \chi_j \nu T \partial_j^2 \right] + \order{\left(\frac{\nu}{D}\right)^2} \; ,
\end{equation}
where
\begin{equation}
    \chi_j \equiv  1-\frac{ \lambda^2 \nu}{D} \int_\mathbb{R} \dslash{q} \frac{q_j^2}{q^\alpha(q^2+r)^2}|V_q|^2  \; .
    \label{eq:chi_j}
\end{equation}
As we had already assumed $V_q$ to be isotropic, then $\chi_j=\chi$ is the same for all the components (\ie, it is independent of the index $j$). It can be easily computed by replacing $q_j^2 \; \rightarrow \; q^2/d$ in the integral, leading to the final result in Eq.~\eqref{eq:finaladiabatic}.

We conclude by noting that our effective equation \eqref{eq:finaladiabatic} is heuristically consistent with the results of Refs.~\cite{theiss1,theiss2}. Similarly, we can consider several copies of the effective Fokker-Planck operator obtained in Ref.~\cite{kaneko} (let us call it $\cor{L}_\T{K}^\T{eff}$), one for each of the field modes $(q,\sigma)$ and particle components $j$. Calling then each of these copies $\cor{L}_\T{K}^\T{eff} \equiv \cor{L}_j^{\sigma,q}$, one can recover our result $\cor{L}^{\T{eff}}$ as 
\begin{eqnarray}
    \cor{L}^{\T{eff}} = \sum_{j=1}^d\sum_{\sigma=R,I} \int \dslash{q}  \cor{L}_j^{\sigma,q} \; .
\end{eqnarray}

\section{Matching between the perturbative and the adiabatic solutions}
\label{par:matching}
In this appendix we investigate how the weak-coupling and the adiabatic approximations for model A dynamics provide the same predictions for $\tau_\phi \ll \tau_X$. We choose for definiteness a Gaussian interacting potential in the form
\begin{equation}
    V_G(\vb{x}) = \frac{1}{(\sqrt{2\pi}R)^d} \exp(-\frac{|\vb{x}|^2}{2R^2}) \; ,
    \label{eq:gaussianpotential}
\end{equation}
where $R$ represents the radius of the colloid, and which reads in Fourier space $V_q = \exp(-q^2 R^2/2)$; the case of the $\delta(\vb{x})$ potential, which models a pointlike colloid, is recovered in the formal limit $R \rightarrow 0$. 

\subsection{Model A with Gaussian interaction potential}
Here we specialize the prediction in Eq.~\eqref{eq:X2_first} for the average particle position to the case of model A dynamics. For simplicity, we rewrite it in the form
\begin{align}
    &\expval*{X_j\t (t)} = \cor{S}_1 + \cor{S}_2 \; ,\\
    &\cor{S}_1  = i \nu D  \int_{0}^t \dd{s_2} e^{-\gamma (t-s_2)} \int_{0}^{s_2} \dd{s_1} e^{-D r (s_2-s_1)} \int \frac{\dd[d]{q}}{(2\pi)^d} q_j e^{-Bq^2-i\vb{C}\cdot\vb{q}} \; ,\\
    &\cor{S}_2  = i \nu^2 T  \int_{0}^t \dd{s_2}   \int_{0}^{s_2} \dd{s_1} e^{-\gamma (t-s_1)} e^{-D r (s_2-s_1)}\int \frac{\dd[d]{q}}{(2\pi)^d} \frac{q_j q^2}{q^2+r}e^{-Bq^2-i\vb{C}\cdot\vb{q}} \; ,
\end{align}
with
\begin{align}
        B &= D(s_2-s_1) + \frac{T}{k}\left[\left(1-e^{-\gamma |s_2-s_1|}\right) - \frac{1}{2}  \left(e^{-\gamma s_1}-e^{-\gamma s_2}\right)^2\right] + R^2 \; , \\
        \vb{C} &=  \left(e^{-\gamma s_1}-e^{-\gamma s_2}\right) \vb{X}_0  \; .
\end{align}
The integration over the momenta $q$ can be performed analytically. Since we have shown in Appendix \ref{par:modelAasymptotics} that $\cor{S}_2$ is subleading for large $t$, here we only report
\begin{equation}
    \cor{S}_1 =i \nu D  \int_{0}^t \dd{s_2} e^{-\gamma (t-s_2)} \int_{0}^{s_2} \dd{s_1} e^{-D r (s_2-s_1)} \int \frac{\dd{\Omega_d}}{(2\pi)^d} \Psi(\Omega) e^{-i C q \Psi(\Omega)} \int_0^\infty \dd{q} q^{d} e^{-Bq^2-i C q \Psi(\Omega)} \; .
\end{equation}
Here we expressed $\vb{C}\cdot\vb{q} = C q \Psi(\Omega)$ in polar coordinates, where $\Psi(\Omega)$ is a suitable director cosine (\eg, $\Psi(\Omega)=1,\,\cos \phi,\, \sin \theta \cos \phi$ for $d=1,2,3$ respectively). Using the properties of Bessel functions \cite{table}, one finds that
\begin{equation}
    \int \frac{\dd{\Omega_d}}{(2\pi)^d} e^{i \vb{q}\cdot \vb{x}} = \frac{J_{d/2-1}(qx) }{(2\pi)^{d/2} (qx)^{d/2-1} } \; ,
    \label{eq:besselprop}
\end{equation}
and since $\Psi(\Omega) \rightarrow - \Psi(\Omega)$ for $\vb{q} \rightarrow -\vb{q}$, we find
\begin{equation}
        \int \frac{\dd{\Omega_d}}{(2\pi)^d} \Psi(\Omega) e^{-i C q \Psi(\Omega)} = - \int \frac{\dd{\Omega_d}}{(2\pi)^d} \Psi(\Omega) e^{i C q \Psi(\Omega)} 
        = \frac{i}{C} \dv{q}    \int \frac{\dd{\Omega_d}}{(2\pi)^d} e^{i \vb{q}\cdot \vb{C}} = \frac{-i J_{d/2}(qC) }{(2\pi)^{d/2} (qC)^{d/2-1} } \; .
\end{equation}
Using the known integral \cite{table}
\begin{equation}
    \int_0^\infty \dd{q} q^{d/2+1} J_{d/2}(qC) e^{-Bq^2} = \frac{C^{d/2}}{\left( 2B \right)^{d/2+1} } e^{-C^2/(4B)} \; ,
\end{equation}
we finally get
\begin{equation}
    \cor{S}_1  = \frac{ \nu D X_0}{\pi^{d/2} 2^{1+d}} e^{-\gamma t } \int_{0}^t \dd{s_2} \int_{0}^{s_2} \dd{s_1} \left[ e^{\gamma (s_2-s_1)}-1 \right]  e^{-D r (s_2-s_1)} \frac{e^{-C^2/(4B)}}{B^{d/2+1} } \; .
    \label{eq:S1_qout}
\end{equation}
Following Sec.~\ref{par:match_main}, we now look for the linear growth coefficient $a$ for large $t$ in the form $\expval*{X\t(t)}e^{\gamma t} \equiv a t$. It is straightforward to derive 
\begin{align}
        a &\equiv \lim_{t \rightarrow \infty} \dv{t} \left[ \expval*{X\t(t)}e^{\gamma t} \right] \n \\
        &= \frac{ \nu D X_0}{\pi^{d/2}\, 2^{1+d}}  \int_{0}^\infty \dd{u} \frac{ \left( e^{-\gamma u}-1 \right)  e^{-D r u}} { \left[ D u + \frac{T}{k}\left(1-e^{-\gamma u}\right) + R^2  \right]^{1+d/2}  } 
        = \frac{ \nu X_0 r^{d/2}}{\pi^{d/2}\, 2^{1+d}}  \int_{0}^\infty \dd{y} \frac{ \left( e^{-\beta y}-1 \right)  e^{-y}} { \left[ y + \frac{rT}{k}\left(1-e^{-\eta y}\right) + R^2 r  \right]^{1+d/2}  } \; ,
    \label{eq:a_linear}
\end{align}
where we have introduced $u=t-s_1$ in the second line and then $y=Dr u$ in the third, calling $\eta \equiv \gamma/(Dr)$ the ratio of the two timescales. Setting $R=0$ in the previous expression we recover the case of the $\delta (x)$ potential; expanding the integral in Eq.~\eqref{eq:a_linear} in powers of $\eta\propto \nu/D$ so as to make contact with the adiabatic approximation, we find
\begin{equation}
    a(R=0) = \frac{\nu \gamma X_0 r^{d/2-1}}{2^{1+d} \pi^{d/2} D  } \Gamma \left(1-d/2 \right) +\order{\eta^2} \; .
    \label{eq:a_R_0}
\end{equation}
Note that this expression is well defined only for $d<2$; this is reminiscent of the fact that the original integrals over $q$ in Eq.~\eqref{eq:X2_first} become UV divergent in $d=2$ if the radius $R$ of the colloid is set equal to zero. By keeping $R$ finite in Eq.~\eqref{eq:a_linear} and introducing the dimensionless variable $\Theta=R^2 r$ we find, instead,
\begin{equation}
    a(R) =  \frac{\nu \gamma X_0 r^{d/2-1}}{2^{1+d} \pi^{d/2} D d}  \left[ e^\Theta \left(2\Theta+d \right) \Gamma\left(1-\frac{d}{2},\Theta\right)-2\Theta^{(2-d)/2} \right] +\order{\eta^2} \; ,
    \label{eq:a_R}
\end{equation}
which is well defined in generic $d$ in terms of the incomplete Gamma function
\begin{equation}
    \Gamma(a,\Theta) = \int_\Theta^\infty \dd{t} t^{a-1} e^{-t} \xrightarrow{\Theta \rightarrow 0} \Gamma(a) \; .
\end{equation}
In $d<2$ we recover Eq.~\eqref{eq:a_R_0} by sending $R\rightarrow 0$ in Eq.~\eqref{eq:a_R}. 

\subsection{Time rescaling factor in the effective Fokker-Planck equation}
The value of $\mu$ defined in Eq.~\eqref{eq:mu} can be easily determined for a rotationally invariant potential $V(x)$ by considering polar coordinates, \ie,
\begin{equation}
        \mu \equiv \frac{\nu}{Dd} \int_\mathbb{R} \dslash{q} \frac{q^{2-\alpha}}{(q^2+r)^2}|V_q|^2 
        = \frac{\nu}{Dd} c_d \int_0^\infty \dd{q} \frac{q^{d+1-\alpha}}{(q^2+r)^2}|V_q|^2 \; ,
    \label{eq:app_mu}
\end{equation}
where the constant $c_d$ was introduced in Eq.~\eqref{eq:solidangle}. We focus here on model A ($\alpha=0$). Choosing $V_q=1$ (point-like colloid), we immediately get from Eq.~\eqref{eq:app_mu} and for $d<2$ 
\begin{equation}
    \mu(R=0) = \frac{\nu r^{d/2-1}}{2^{1+d} \pi^{d/2} D } \Gamma \left(1-\frac{d}{2} \right) \; ,
    \label{eq:mu_R_0}
\end{equation}
where we used the relation $\sin(\pi x)\Gamma(x) \Gamma(1-x) = \pi$ \cite{arfken}. By choosing instead a Gaussian potential $V_q=\exp(-q^2R^2/2)$, we can use the representation \cite{arfken}
\begin{equation}
    A^{-n} = \frac{1}{\Gamma(n)} \int_0^\infty \dd{y} y^{n-1} e^{-A y}
\end{equation}
to express
\begin{equation}
    \mu = \frac{\nu c_d}{Dd} \int_0^\infty \dd{y} y e^{-y r} \int_0^\infty \dd{q} q^{d+1} e^{-q^2(R^2+y)}
    =  \frac{\nu r^{d/2-1}}{2^{1+d} \pi^{d/2} Dd}  \left[ e^\Theta \left(2\Theta+d \right) \Gamma\left(1-\frac{d}{2},\Theta\right)-2\Theta^{(2-d)/2} \right] \; ,
    \label{eq:mu_R}
\end{equation}
where again we called $\Theta=R^2 r$. Comparing Eqs.~\eqref{eq:a_R_0} and \eqref{eq:a_R} with Eqs.~\eqref{eq:mu_R_0} and \eqref{eq:mu_R}, one can note that $a= \gamma X_0 \mu$ to the perturbative order at which we are working. We can thus conclude that the weak-coupling and the adiabatic approximations provide the same results in the case of model A field dynamics whenever $Dr\gg \gamma$, and they match according to Eq.~\eqref{eq:balancing}. 

On the other hand, we have seen that they no longer agree when $Dr\sim \gamma$ (and in particular when the field is at criticality), and that no agreement is generically found, as expected, in the case of model B field dynamics. 

\section{Numerical simulation}
\label{par:appendix_numerical}
Numerical simulations are performed by direct integration of the coupled Langevin equations of motion \eqref{eq:field} and \eqref{eq:particle}. Field variables are discretized as $\{ \phi_i(t) \}_{i=1}^N$ with $\phi_i(t)\equiv \phi(\vb{x}_i,t) \in \mathbb{R}$, and they sit on the $N=L^d$ sites of a $d$-dimensional hypercubic lattice of side length $L$. Distances are measured in units of the lattice spacing $a$, which we retain for clarity in the following formulas, but which will be eventually set to unity. The coordinate $\vb{X}(t) \in \mathbb{R}^d$ of the \textit{center} of the particle is taken to be real-valued, \ie, the particle is not constrained to move on the lattice sites only. Upon integration by parts, the equation of motion of the particle can be rewritten as
\begin{align}
        \dot{\vb{X}}(t) & = -\nu k\vb{X} + \nu\lambda \int \dd[d]{x} V(\vb{x}-\vb{X}) \grad \phi(\vb{x}) + \bm{\xi}(t) \n \\
        &\simeq -\nu k\vb{X} + \nu\lambda \sum_{i=1}^N V(\vb{x}_i-\vb{X}) \widetilde{\grad} \phi_i + \bm{\xi}(t) \; ,
    \label{eq:num_particle}
\end{align}
where we introduced the discrete gradient
\begin{equation}
    \widetilde{\grad}_j \phi_i = \frac{ \phi(\vb{x}_i+\hat{\bm{\mu}}_j)-\phi(\vb{x}_i-\hat{\bm{\mu}}_j)}{2a} \; ,
\end{equation}
with $\hat{\bm{\mu}}_j$ locating the position of the $2$ neighbouring sites of each $\vb{x}_i$ along direction $j$.
The discretized equation of motion for the field $\phi_i$ in model A reads
\begin{equation}
    \partial_t\phi_i(t)= -D \left[(r-\widetilde{\Delta})\phi_i(t)-\lambda V(\vb{x}_i-\vb{X}(t)) \right] + \zeta_i(t) \; ,
    \label{eq:num_A}
\end{equation}
where $\zeta_i(t)$ is a Gaussian random variable with variance $\expval*{\zeta_i(t)\zeta_j(t')}= 2DT a^{-1} \delta_{ij}\delta(t-t')$. We also defined the discrete Laplacian
\begin{equation}
    \widetilde{\Delta} \phi_i = \frac{1}{a^2}  \sum_{\langle k,i \rangle} \left( \phi_k - \phi_i \right) \; ,
\end{equation}
where the sum runs over the $2d$ neighbouring sites of $\vb{x}_i$. Similarly, the discretized equation of motion for the field in model B reads
\begin{equation}
        \partial_t\phi_i(t)= D \widetilde{\Delta} \left[(r- \widetilde{\Delta} )\phi_i(t)-\lambda V(\vb{x}_i-\vb{X}(t)) \right] + \widetilde{\grad} \cdot \bm{\eta}_i(t)
    \label{eq:num_B}
\end{equation}
where $\bm{\eta}_i(t)$ is a vectorial noise with zero mean and variance $\expval*{\eta_i^{(\alpha)}(t)\eta_j^{(\beta)}(t')}= 2DT a^{-1} \delta_{ij}\delta_{\alpha \beta}\delta(t-t')$, and we take its discrete divergence as $\widetilde{\grad}_\alpha \eta_i^{(\alpha)}(t)$. We chose in both cases a Gaussian interaction potential as in Eq.~\eqref{eq:gaussianpotential}, which yields a smooth expression for its Laplacian
\begin{equation}
    \laplacian V_\T{G}(\vb{x}) = \frac{\abs{\vb{x}}^2-R^2 d}{R^4} V_\T{G}(\vb{x}) \; .
\end{equation}
Equations \eqref{eq:num_particle} and \eqref{eq:num_A} (or \eqref{eq:num_B}) represent a set of $(N+d)$ ordinary stochastic differential equations which can be integrated by standard methods in real space \cite{frenkel}. We choose a simple Euler-Maruyama scheme (order $\Delta t^{1/2}$) for the evolution of the field variables and a more refined method, \ie, the Stochastic Runge-Kutta (order $\Delta t^{3/2}$, see Ref.~\cite{stochasticRK}), for the particle coordinate. We expect this to improve the stability of the particle dynamics in spite of the lower-order algorithm adopted for the field, because the latter only contributes at $\order{\lambda\ll 1}$ to the evolution of the particle. 

At the beginning of each trial, we prepare the field in its equilibrium distribution at temperature $T$ in Fourier space and then move back to real space using a discrete Fourier transform. We then add the colloid at position $X_0\neq 0$, and record its relaxation trajectory as it moves towards the center of the harmonic trap. Simulations performed at temperature $T\sim \order{10^{-1}}$ on a lattice with side $L\sim \order{10^3}$, such as the one shown in Fig.~\ref{fig:fits}, require $\order{10^8-10^9}$ trials in order to obtain a clear sample of the algebraic decay of the average particle position. Indeed, the signal/noise ratio becomes increasingly small at long times, which is the region we are mostly interested in.

The complete code written in C is made available open source on GitHub \cite{code}.

\section{A first quantitative estimate}
\label{app:relevance}
In this Appendix we attempt a comparison between our model with off-critical model B dynamics, and experiments performed on colloidal particles in binary liquid mixtures. Even though our model is not meant to give a realistic description of such a physical system (for instance, hydrodynamic effects are ignored), it is still interesting to inspect the typical orders of magnitude and check how large the algebraic behavior of the average particle position can be made, compared with its radius $R$.

Following Sections \ref{par:transient} and \ref{par:relevance}, we start by choosing the value of the initial displacement $X_0$ in order to maximize the amplitude of the particle position at the crossover time $t_c$. This was given in Eq.~\eqref{eq:crossover_amplitude_dimensionless}, which suggests to take $X_0$ as small as possible, but still sufficiently large so that the assumption $t_c > \tau_X$ we made in Section \ref{par:transient} is still satisfied. Recall that $t_c$ is the time taken by the field in order to relax over a length scale $\sim X_0$, and it can be identified in the off-critical model B with $t_c\sim X_0^2/Dr$.

What is the typical size of $t_c$? While $r=\xi^{-1/2}$ and it is simple to plug in typical values for the correlation length $\xi$ which can be obtained in experiments, it is not obvious how large a realistic $D$ is. Within model B, we learn from Eq.~\eqref{eq:tau_phi} that $\tau_\phi^{-1} \simeq Dr q^2$ for wavelengths $q \ll r^{1/2} = 1/\xi$. However, real binary fluid mixtures are generally described by model H \cite{halperin,Tauber}, where the field relaxation time for $q\xi \ll 1$ is given within mode-coupling theory by $\tau_\phi^{-1} \simeq D_\xi q^2$, with \cite{onuki}
\begin{equation}
    D_\xi = \frac{k_B T}{6 \pi \eta \xi} \; ,
\end{equation}
being $\eta$ the fluid viscosity. Notice the similarity with the free diffusion coefficient of the colloid $\expval{X^2(t)}\simeq D_R t$,
\begin{equation}
    D_R = \frac{k_B T}{6 \pi \eta R} \; .
\end{equation}
Typical colloid radii are of the order of $R\simeq 1\mu$m, while typical correlation lengths obtainable with a water-lutidine mixture are or the order of a few tens of nanometers \cite{casimirColloids,energytransfer}. To give a heuristic estimate of $D$, we compare the diffusion coefficient of the order parameter fluctuations in model B with that of model H, thus identifying $Dr \simeq D_\xi$, which renders $t_c \simeq X_0^2/D_\xi$.

Equation \eqref{eq:crossover_amplitude_dimensionless} still contains the dimensionless parameter $g$, which sets the strength of the interaction between the field and the particle. Its amplitude will depend on the specific coupling mechanism realized in a certain experiment, and clearly the overall effect will be enhanced if $g$ can be made larger. However, here we take $g\sim 1$ in order to remain within the perturbative regime, under which most of the analyses in this work were carried out.

To fix the ideas, we take $t_c \sim 4 \tau_X$, whence $X_0 \simeq \sqrt{2 \tau_X  D_\xi}$. From Eq.~\eqref{eq:crossover_amplitude_dimensionless} we read
\begin{equation}
    \frac{\expval*{X_j (t_c)}}{R} \sim \left( \frac{4  \tau_X D_\xi }{R^2}\right)^{(1-d-z)/2} \sim \left( \frac{200 \tau_X D_R}{R^2}\right)^{-(d+3)/2} ,
\end{equation}
where in the last passage we inserted the realistic estimate $\xi\sim R/50$ and we set $z=4$. Now we notice that $\tau_d \sim R^2/D_R $ is the timescale of thermal diffusion of the colloid over a distance of the order of its own radius. A typical value for the free diffusion coefficent is $D_R \simeq 0.22 (\mu m)^2s^{-1}$ \cite{casimirColloids}, whence $\tau_d \sim 4-5$s. We can conclude that
\begin{equation}
    \frac{\expval*{X_j (t_c)}}{R} \sim \left( \frac{\tau_d }{200 \tau_X}\right)^{(d+3)/2} \; .
\end{equation}
Typical timescales $\tau_X$ of relaxation of colloidal particles trapped by optical tweezers are of the order of a few tens of milliseconds \cite{energytransfer}. It then appears that $\expval{X(t_c)}$ measured in units of the colloid radius $R$ can be made as large as $10^{-1}$ at least. It should be stressed that digital video-microscopic observation of $2\mu$m-sized silica particles immersed in binary liquid mixtures currently allows to resolve displacements of up to 5nm. We are thus led to conclude that, even if $\expval{X(t_c)}$ is indeed small compared to the colloid radius, the effect we predicted could still be detected experimentally.
\end{document}